\documentclass[12pt]{article}
\usepackage[utf8]{inputenc}
\usepackage{graphicx}
\usepackage{amsmath}
\usepackage{amssymb}
\usepackage{bm}
\usepackage{booktabs}
\usepackage{url}
\usepackage{tikz}
\usetikzlibrary{positioning}
\usepackage{authblk}
\usepackage{svg}
\usepackage{microtype}
\usepackage[hidelinks]{hyperref}
\hypersetup{
    pdftitle={Statistical Modelling of Planetary Boundary Layer Height and Its Uncertainty Using GRUAN Profiles},
    pdfauthor={Tommaso Locatelli},
    pdfkeywords={Planetary Boundary Layer, GRUAN, Statistical Modelling},
    pdfsubject={Research Paper},
    colorlinks=true,
    linkcolor=blue,
    citecolor=blue,
    urlcolor=blue
}

\title{Statistical Modelling of Planetary Boundary Layer Height and Its Measurement Uncertainty Using GRUAN Profiles}

\author[1,2,3]{Tommaso Locatelli}
\author[3]{Alessandro Fassò}
\author[4]{Fabio Madonna}

\affil[1]{University School for Advanced Studies IUSS Pavia}
\affil[2]{Institute of Integrated Methodologies for Earth Observation, CNR-IMIOT}
\affil[3]{University of Bergamo}
\affil[4]{University of Salerno}

\date{July 2026}

\begin{document}

\maketitle

\begin{abstract}
The Planetary Boundary Layer (PBL) governs the exchange of energy and moisture and hosts the highest concentrations of pollutants before they mix into the free troposphere. The height of the PBL (PBLH) is therefore a key variable in meteorological and air‑quality applications. Despite the wide range of methods available to derive PBLH from atmospheric observations, the associated uncertainties are rarely quantified. This study presents a methodology for propagating radiosonde measurement uncertainty into PBLH estimates obtained from state‑of‑the‑art retrieval methods, including the parcel method, gradient‑based methods, and the Richardson‑number method. The framework relies on three components. First, it uses the GCOS Reference Upper‑Air Network (GRUAN) Data Product, which provides traceable uncertainty estimates for all variables required in PBLH retrievals. Second, it employs a state‑space model that captures the structure of atmospheric profiles and enables the generation of physically plausible simulated vertical profiles consistent with both observations and their uncertainties. Third, a Monte Carlo approach is used to propagate measurement uncertainty into the PBLH estimates, refining the retrieval and quantifying its uncertainty. Beyond providing uncertainty estimates, the methodology also shows preliminary signs of increased robustness in PBLH detection across several case studies, particularly in situations where standard gradient‑based methods exhibit sensitivity to measurement uncertainty.
\end{abstract}

\section{Introduction}
The Planetary Boundary Layer (PBL), the lowest part of the atmosphere, governs the exchange of energy and moisture and is the zone where pollutants reach their highest concentrations before entering the free troposphere \cite{Stull1988,Seibert2000PBLH}. It acts as the primary zone where pollutants and water vapour are trapped, reaching high concentrations \cite{Holzworth1967}. Conversely, the PBL is also subject to entrainment processes, with clean air or aged pollutants intruding from the free troposphere, altering local air quality \cite{Stevens2002}. Within this dynamic environment, water vapor plays a central thermodynamic role, acting as both the dominant greenhouse gas and the primary vehicle for latent heat transport \cite{WallaceHobbs2006}. Under this mechanism, in cloud-topped conditions, the thermal energy stored during surface evaporation is later released, directly driving buoyant convective plumes that regulate the development and diurnal growth of the PBL height (PBLH) \cite{Stull1988}.

Furthermore, the PBL is the critical domain for wind energy harvesting, being mechanical energy extracted from the atmospheric flow \cite{Burton2011}. PBLH is therefore a key variable in meteorology, renewable energies \cite{wes-10-1681-2025}, and air‑quality applications \cite{Seibert2000PBLH}. Long-term monitoring of PBLH trends is also relevant for understanding climate change impacts and variations in the atmospheric pollutant dilution capacity over the years. 

During convective daytime conditions, the PBL is generally dominated by intense thermal buoyancy, forming a fully convective state. In these conditions, the PBL is defined as the Mixing Layer. The Mixing Layer Height (MLH) represents the vertical extent to which gaseous pollutants, aerosols, and thermal energy are mixed vigorously and homogeneously by turbulent eddies. The MLH determines the available dilution volume for surface emissions, directly regulating ground-level pollutant concentrations and smog formation \cite{Holzworth1967,WallaceHobbs2006}.

At the sunset transition, the boundary layer dynamics become highly complex; as solar forcing diminishes, convective turbulence rapidly decays, causing the daytime Mixing Layer to collapse. This thermodynamic shutdown leaves a neutrally stratified Residual Layer aloft, which effectively decouples from the surface physics \cite{GrimsdellAngevine2002}. 

Subsequently, during the night, the radiative cooling of the Earth's surface dampens convective turbulence, transforming the atmosphere into a stably stratified regime, forming the Nocturnal Boundary Layer (NBL). Detecting the NBLH under these conditions is  more challenging. The NBL is typically very shallow, often capped by a low-level jet or characterized by intermittent, localized turbulence that standard remote sensing instruments fail to resolve properly due to sampling issues, such as their blind zones or limited vertical resolution \cite{LiuLiang2010}.

Exisiting PBLH detection algorithms are based on a heterogeneous array of observation systems, including radiosondes, lidar and ceilometer backscatter measurements, microwave radiometers, radar wind profilers \cite{Summa2022HyMeXABL}, and satellite‑based retrievals such as GNSS‑RO \cite{Mannucci}. Furthermore, different evaluation criteria, known as PBLH methods, are applied to meteorological variables and atmospheric properties, such as virtual potential temperature, humidity gradients, aerosol backscatter, refractivity, and turbulence metrics, leading to significant dispersion in the retrieved PBL heights \cite{seidel_estimating_2010}.

These methodological discrepancies further amplify the overall uncertainty. Parametric uncertainty arises from the choice of diagnostic criteria and thermodynamic tracers, while sampling uncertainty is related to partial and non-uniform observation of the atmospheric state, both of which can substantially affect PBLH estimates \cite{Madonna2021ABLTrends}. Moreover, the propagation of  structural and  sampling uncertainties in climatological assessments has also been investigated \cite{seidel_estimating_2010}.

Addressing structural and sampling uncertainties in its calculation is therefore essential for improving the parametrization of vertical transport in numerical weather prediction \cite{Zhao2021} and chemical transport models \cite{KIM2024176838}.

Within this context, the atmospheric research community has increasingly turned to ensemble approaches, i.e.. using different algoorithms and use their mean and the best estimate, to mitigate the long-standing challenges of gradient attribution and structural uncertainties.  Establishing a robust uncertainty framework is further complicated by the fact that an absolute reference or ground truth does not exist for the PBLH. In this context, while ensemble-based approaches (\cite{Chen2023PBLHEnsemble}, \cite{amt-17-3029-2024} ) to produce a best-estimate, either from measurement or models (\cite{cds_era5_catalogue}) are primarily used to reduce  structural uncertainties and attribution ambiguities, they do not involved the measurement uncertainties.

As an indirect quantity, the PBLH is also affected by the measurement uncertainty present in the input variables from which it is derived. However, the PBLH uncertainty associated with the propagation of measurement errors affecting a specific atmospheric variable has not yet been explicitly addressed in the literature. That is, when the observation system, the evaluation method, and the spatiotemporal context are fixed, the uncertainty arising solely from the measurement itself remains unquantified. Quantifying measurement uncertainty through its explicit propagation along a vertical atmospheric profile represents a paradigm shift, enabling a physically consistent definition of the true uncertainty of the boundary‑layer height. A proper treatment of measurement uncertainty not only improves the reliability of PBLH estimates but also strengthens the overall methodology by making it less prone to measurement noise and variability in the input measurements.

This paper advances current methodologies by embedding established PBLH detection techniques within a rigorous state‑space modeling framework and by explicitly incorporating the measurement uncertainty provided by the GRUAN reference observing network, thereby enabling a revised application of PBLH diagnostics under quantified observational uncertainty.

The structure of the paper is as follows. Section~\ref{sec:2} introduces the radiosonde observations used for the PBLH estimation and the GRUAN Data Product (GDP) employed in this study as reference measurements with resolved vertical uncertainty. Section~\ref{sec:3} explains the proposed methodology by introducing the formalization of the PBLH methods, describing the standard plug‑in estimation approach, presenting the state‑space model and the subsequent PBLH estimation and outlining the Monte Carlo–based estimation of PBLH and its uncertainty via simulation smoothing. Section~\ref{sec:4} provides illustrative examples that clarify the application of the methodology, followed by an aggregated analysis. Finally, Section~\ref{sec:5} discusses the main findings and proposes future
directions for validation and further development. Three supplementary sections
report the code availability (Section~\ref{sec:code}), data availability
(Section~\ref{sec:data}), and AI usage (Section~\ref{sec:AI}). The list of tables, together with the tables complementary to the aggregated 
analysis, is provided after the references in Section~\ref{sec:tbls}.

\section{Radiosonde Data and GRUAN Dataset}
\label{sec:2}

Among the various observing systems, radiosonde (RS) profiles have been extensively used in PBLH estimate, both in early climatological studies and in the evaluation of retrieval algorithms from other remote sensing instrumens \cite{essd-16-1-2024}, owing to their high vertical resolution and long-term global record \cite{amt-14-5977-2021}.

Balloon‑borne radiosonde observations provide in‑situ measurements of temperature, humidity, pressure, and wind—often with high vertical resolution—and constitute the historical reference system for atmospheric profiling used in global meteorological networks since the mid‑20th century \cite{VOMEL202323}.

Despite their role as the reference system for upper‑air profiling, radiosonde observations are limited by sparse global coverage, heterogeneous station distribution, and coarse temporal sampling, which prevent them from resolving the full diurnal cycle of the PBLH \cite{essd-16-1-2024}.

This study uses the GRUAN Data Products (GDPs) RS41-GDP.1, which provides vertically resolved upper-air measurements traceable to the SI and with quantified measurement uncertainties \cite{GRUAN-TD-8}, enabling a rigorous treatment of uncertainty propagation in the estimation of any further products from the vertical profiles.

The Global Climate Observing System (GCOS) Reference Upper‑Air Network (GRUAN) is an international reference network designed to fill a critical gap in the global atmospheric observing system. GRUAN provides long‑term, high‑quality climate data records from the surface through the troposphere and into the stratosphere, based on rigorously characterized measurements of essential climate variables. These reference observations are used to determine atmospheric trends, constrain and calibrate more spatially comprehensive observing systems, such as satellites and operational radiosonde networks. GRUAN is a global network of 35 sites, 16 of which have been certified \cite{GRUAN_2026}.

\begin{figure}[!h]
    \centering
    \includegraphics[width=1\linewidth]{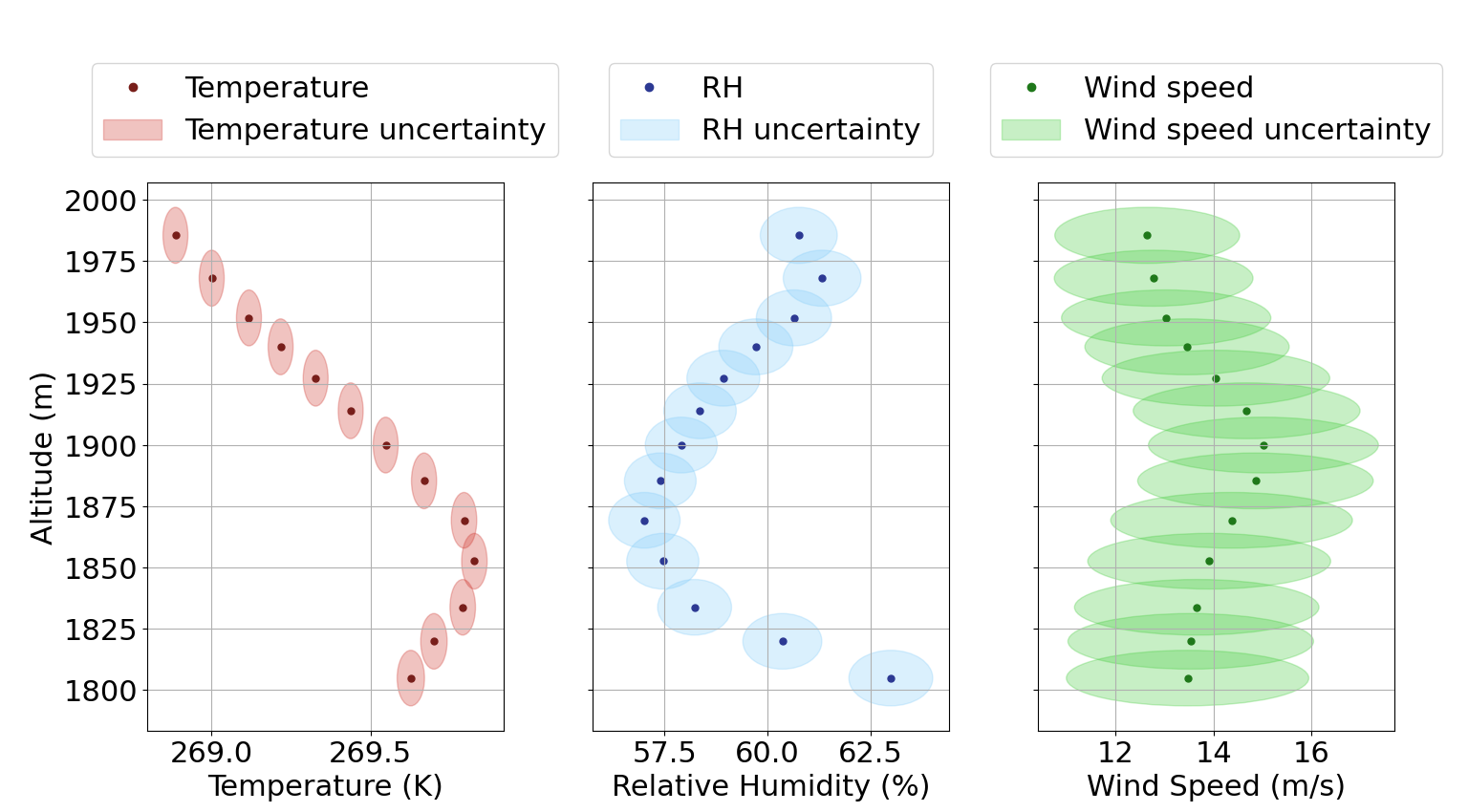}
    \caption{Profile of GRUAN RS41-GDP.1 sounding from the Lindenberg site. Temperature, relative humidity, and wind speed are shown between 1800 and 2000\,m, together with their vertically resolved uncertainty ellipses. Dots represent the measured variables, while the shaded ellipses illustrate the expanded uncertainties ($k=2$) in both the horizontal variable dimension (ellipse width) and the altitude dimension (ellipse height).}

    \label{fig:gdp_unc}
\end{figure}

Figure \ref{fig:gdp_unc} shows an example of GDP profile from the Lindenberg site, including temperature (K), relative humidity (\%) and wind speed (m/s) as a function of altitude above sea level (m), together with their expanded uncertainties, computed with a coverage factor $k=2$. Note that the altitude itself is also affected by measurement uncertainty. Given a generic variable $x$, we denote the standard uncertainty of $x$ by $\sigma_x$, and the expanded uncertainty by $u(x) = k\,\sigma_x$.

For the scope of this research, three GRUAN sites were selected to represent different climatic regions: Lindenberg (LIN), Hong Kong (HKO), and Lauder (LAU). All the data for 2024 are considered. The selected stations performed at least two soundings per day. Lindenberg typically launches four radiosondes per day, so only the midnight (00:00 UTC) and noon (12:00 UTC) soundings are retained for consistency across sites.

Table \ref{tbl:sites} provides detailed location information for the selected sites.

\begin{table}[h!]
\centering
\caption{Geographical information of the sites.}
\label{tbl:sites}
\begin{tabular}{l l c c c}
\hline
\textbf{Site} & \textbf{Country} & \textbf{Latitude} & \textbf{Longitude} & \textbf{Altitude} \\
\hline
LIN (Lindenberg) & Germany & 52.21°N & 14.12°E & 98 m \\
HKG (Hong Kong) & China & 22.31°N & 114.17°E & 65 m \\
LAU (Lauder) & New Zealand & 45.05°S & 169.67°E & 3 m \\
\hline
\end{tabular}
\end{table}

\begin{figure}[!h]
    \centering
    \includegraphics[width=\linewidth]{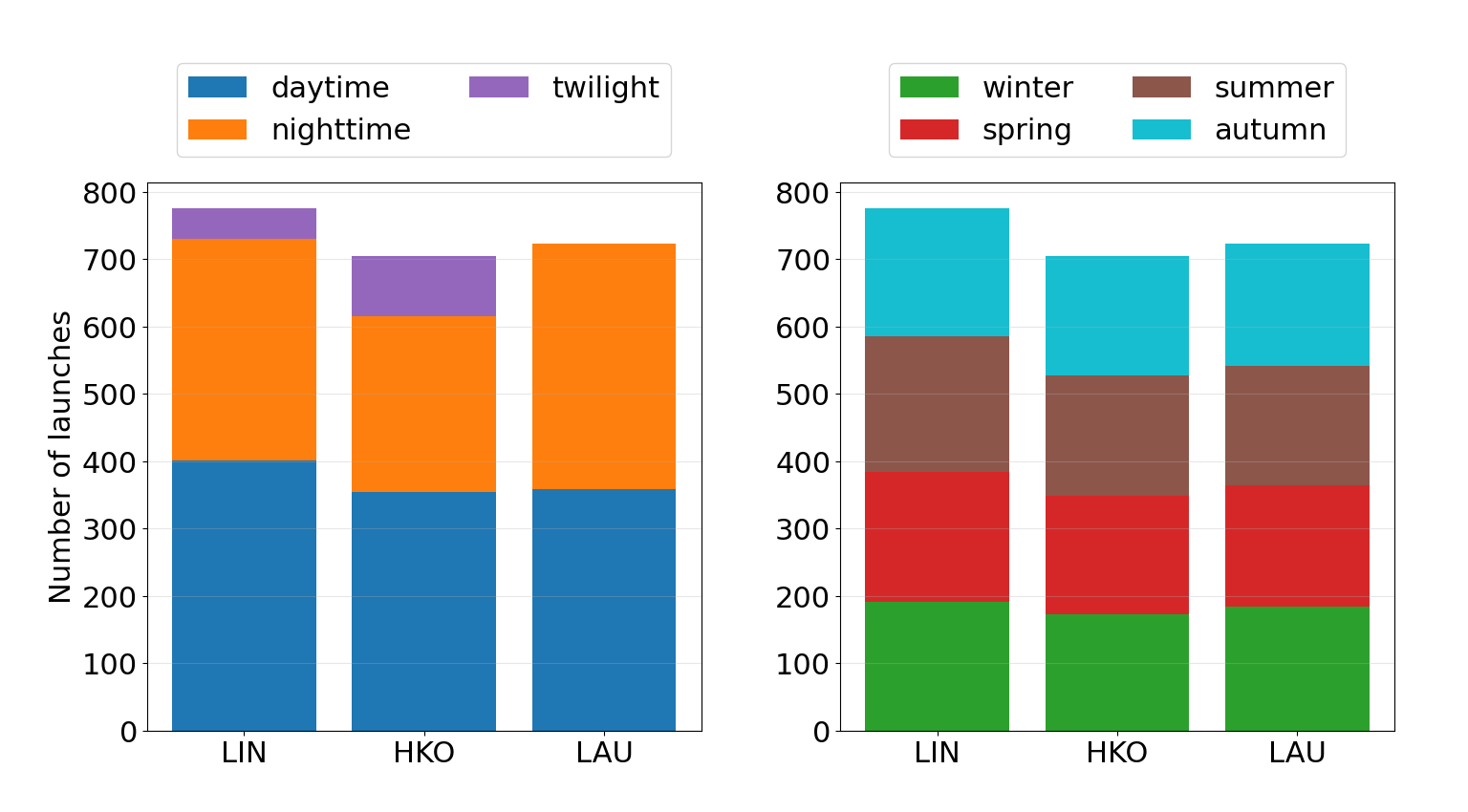}
    \caption{Day/night/twilight (left) and seasonal (right) distribution of radiosonde launches at LIN, HKO, and LAU during 2024. Classifications follow GRUAN metadata and local-time season definitions.}
    \label{fig:dataset_distr}
\end{figure}

Figure \ref{fig:dataset_distr} shows the distribution of radiosonde launches across the three GRUAN sites, separated into daytime, nighttime, and twilight categories provided by the GDP and computed from the solar elevation angle (left panel), together with the corresponding seasonal distribution in local time (right panel).

\section{Methodology}
\label{sec:3}

This section introduces a formal description of four PBLH methods, discusses their 
standard plug-in application to uncertainty-affected measurements, revises them 
within a state-space model framework \cite{durbin}, and exploits this latter 
representation to obtain a Monte Carlo procedure for the estimation and uncertainty 
quantification of the PBLH.

For this purpose, four PBLH methods have been selected to cover different variables and physical processes relevant to PBLH detection, although the framework can be easily extended to additional approaches. The selected methods are: 
\begin{itemize}
    \item the parcel method (PM), which retrieves the mixing height representing the level at which an air parcel becomes neutrally buoyant with respect to vertical motion \cite{Holzworth1967};
    \item the vertical gradient of virtual potential temperature method ($\theta_v$), which identifies the maximum gradient in virtual potential temperature associated with the capping inversion isolating the PBL air mass \cite{Stull1988};
    \item the vertical gradient of relative humidity method (RH), which detects the sharp decrease in humidity marking the transition from the moist PBL to the drier free troposphere \cite{seidel_estimating_2010};
    \item the Richardson number method ($RM$), which identifies the change from a turbulent PBL regime to a more laminar free–tropospheric flow \cite{seidel2012}.
\end{itemize}

\subsection{PBLH Methods}
\label{sec:3.1}

Let $t = 1,\dots,n$ denote the time, expressed in seconds, from the beginning of the radiosonde flight to the ballon burst.

Consider the thermodynamic variables along the radiosonde (RS) flight, collected in a single vector at any time $t$, denoted by
\[
\mathbf{x}_{t} = (z_t,\, T_t,\, p_t,\, RH_t,\, r_t,\, u_t,\, v_t)'
\]
where $z_t$ is the altitude above sea level (m), $T_t$ is temperature (K), $p_t$ is pressure (hPa), $RH_t$ is relative humidity (\%), $r_t$ is the water-vapour mass mixing ratio (kg\,kg$^{-1}$), and $u_t$ and $v_t$ are the zonal and meridional wind components (m\,s$^{-1}$). We denote by $X_n$ the set of all $\mathbf{x}_t$ for $t = 1,\dots,n$.

Based on such a vector, let us define the diagnostic variables vector
as a vector indexed by $t$ containing the variables that are not 
measured directly, but on which the criteria of PBLH methods are applied. 
Formally,
\[
\mathbf{y}_{t} = f(\mathbf{x}_{t}) 
= (\theta_{v,t},\, \Gamma_{\theta_v,t},\, \Gamma_{RH,t},\, Ri_{t})',
\]
with $\theta_{v,t}$ virtual potential temperature (K), $\Gamma_{\theta_v,t}$ virtual potential-temperature vertical 
gradient (K/m), $\Gamma_{\theta_v,t}$ relative-humidity vertical gradient (\%/m), and $Ri_{t}$ Richardson number (pure number), defined by
\[
\theta_{v,t}
= T_{t} \left( \frac{p_0}{p_{t}} \right)^{\kappa} 
  \left( 1 + 0.61\, r_{t} \right),
\]
\[
\Gamma_{\theta_v,t}
= \frac{\partial \theta_{v,t}}{\partial z},
\]
\[
\Gamma_{RH,t}
= \frac{\partial RH_{t}}{\partial z},
\]
\[
Ri_{t}
= \frac{\left(\frac{g}{\theta_{v,t}}\right)
        \left(\frac{\partial \theta_{v,t}}{\partial z}\right)}
       {\left(\frac{\partial u_{t}}{\partial z}\right)^2
      + \left(\frac{\partial v_{t}}{\partial z}\right)^2}.
\]
We also denote the entire vertical profile of $\mathbf{y}_{t}$ by $Y_n=\{\mathbf{y}_{t}:t=1,\dots,n\}$.
Note that the function $f$ has the role of a measurement function in the 
mathematical model for indirect measurements within the 
GUM framework \cite{GUM2008}.

Then, let define the PBLH vector containing the PBLH according to a list of PBLH methods
 at a specific time. Formally,
\[
\mathbf{h}=g(Y_n)=(h_{PM},\,h_{\theta_v},\,h_{RH},\,h_{RM})'
\]
where four methods can be formalized as follows;
\begin{itemize}
    \item Parcel Method: $h_{PM}=\min_{z_t}\{z_t:\theta_{v,t}\geq\theta_{v,1}\}$,
    \item Theta Method: $h_{\theta_v}=\arg\max_{z_t}\{\Gamma_{\theta_v,t}\}$
    \item RH Method: $h_{RH}=\arg\min_{z_t}\{\Gamma_{RH,t}\}$
    \item Richardson Method: $h_{RM}=\min_{z_t}\{z_t:Ri_{t}\geq 0.25\}$.
\end{itemize}
with $\theta_{v,1}$ denotes the virtual potential temperature at the surface, being $t=1$ the first measuring istant.

Note that the function $g$ simply applies the corresponding PBLH criteria, in a manner analogous to an optimization procedure, to the vertical profile of the diagnostic-variables vector. These quantities serve a theoretical purpose, as the measurements are affected by uncertainty; that is, each $\mathbf{x}_{t}$ is affected by some measurement error. The next section (\ref{sec:3.2}) illustrates how standard PBLH estimates are typically obtained through a plug-in approach.

\subsection{Standard Plug-in Estimation}
\label{sec:3.2}

Let the observed vector, affected by measurement uncertainty, be denoted by
\[
\hat{\mathbf{x}}_t = (\hat z_t,\, \hat T_t,\, \hat p_t,\, \hat{RH}_t,\, 
\hat r_t,\, \hat u_t,\, \hat v_t)' .
\]
Analogously, $\hat{X}_n=\{\hat{\mathbf{x}}_t:t=1,\dots,n\}$. The hat symbol indicates that the observations are affected by 
measurement uncertainties. We denote the corresponding vector of 
standard measurement uncertainties, provided by GDPs, as
\[
\boldsymbol{\sigma}_{\hat{\mathbf{x}},t}
=(\sigma_{\hat{z},t},\,\sigma_{\hat{T},t},\,\sigma_{\hat{p},t},\,\sigma_{\hat{RH},t},\,\sigma_{\hat{r},t},\,\sigma_{\hat{u},t},\,\sigma_{\hat{v},t})'.
\]
The standard plug‑in implementation of PBLH methods based on $\hat{X}_n$ assumes some approximations in order to obtain an estimate of the diagnostic‑variables vector. Formally,
\[
\hat{\mathbf{y}}_t=\hat{f_t}(\hat{X}_n)
=(\hat{\theta}_{v,t},\,\hat{\Gamma}_{\theta_v,t},\,\hat{\Gamma}_{RH,t},\,\hat{Ri}_t)',
\]
where the observed values are used as proxies for the true atmospheric state, gradients are estimated by finite‑difference ratios, and the bulk Richardson number is used:
\[
\hat{\theta}_{v,t}
=\hat T_{t} \left( \frac{p_0}{\hat p_{t}} \right)^{\kappa} 
  \left( 1 + 0.61\, \hat r_{t} \right),
\]
\[
\hat{\Gamma}_{\theta_v,t}
=\frac{\Delta\hat{\theta}_{v,t}}{\Delta \hat z_t}
=\frac{\hat{\theta}_{v,t}-\hat{\theta}_{v,t-1}}{\hat z_t-\hat z_{t-1}},
\]
\[
\hat{\Gamma}_{RH,t}
=\frac{\Delta \hat{RH}_t}{\Delta \hat z_t}
=\frac{\hat{RH}_t-\hat{RH}_{t-1}}{\hat{z}_t-\hat{z}_{t-1}},
\]
\[
\hat{Ri}_t
=\frac{({g}/{\hat{\theta}_{v,t}})
      (\hat{\theta}_{v,t}-\hat{\theta}_{v,1})(\hat z_t-\hat z_1)}
      {(\hat u_t-\hat u_1)^2+(\hat v_t-\hat v_1)^2}.
\]
Based on this approximation, it is possible to estimate the associated 
uncertainty by applying the error propagation law, where the 
vector $\boldsymbol{\sigma}_{\mathbf{x},t}$ provides the input‑quantity uncertainties. This yields
\[
\boldsymbol{\sigma}_{\hat{\mathbf{y}},t}
=(\sigma_{\hat{\theta}_v,t},\,\sigma_{\hat{\Gamma}_{\theta_v},t},\,\sigma_{\hat{\Gamma}_{RH},t},\,\sigma_{\hat{Ri},t})'.
\]

Finally, from the estimated diagnostic-variables vectors it is 
possible to obtain the estimates of the PBLH identified by the PBLH methods 
of interest simply by applying the criterion function $g$ to 
$\hat{Y}_n=\{\hat{\mathbf{y}}_t:t=1,\dots,n\}$. Formally,
\begin{equation}
\hat{\mathbf{h}}
=
g\bigl(\hat{Y}_n)
=
(\hat h_{PM},\, \hat h_{\theta_v},\, \hat h_{RH},\, \hat h_{RM})' .
\label{eq:plug-in_estimate}
\end{equation}
Contrary to the previous diagnostic step, it is not possible to propagate 
measurement uncertainty to $\hat{\mathbf{h}}$ using the law of propagation 
of uncertainty, because the derivatives required for the function $g$ cannot be calculated analytically. Therefore, although $\boldsymbol{\sigma}_{\hat{\mathbf{y}},t}$ provides an 
indication of the uncertainty associated with the standard estimation 
process, no state-of-the-art PBLH method currently provides an estimate 
for $\boldsymbol{\sigma}_{\hat{\mathbf{h}}}$.

How to refine these methods, account for the measurement uncertainty and obtain a meaningful propagation of measurement uncertainty to will be explained in the next two  (\ref{sec:3.3}, \ref{sec:3.4}).

\subsection{State-Space Model Estimation}
\label{sec:3.3}

The state-space model approach is based on the idea that the dynamics of a system evolve in a space of state vectors, while these dynamics are indirectly observed through a transformation of the state vectors, namely the observation vector. A state equation defines the system dynamics, whereas a measurement equation expresses how observations are obtained. This approach enables 
measurement uncertainty to be directly embedded in the model formulation. Several techniques have been developed to retrieve estimates of the unobserved state \cite{durbin}. In this study, a simple but effective state-space specification, known as the Local Linear Trend model, is employed to obtain robust estimates of variable gradients in the presence of measurement uncertainty \cite{shumstof2000}.

First, let us redefine the observation vector in a more convenient 
form as:
\[
\bar{\mathbf{x}}_t = (\hat z_t,\, \hat{\theta}_{v,t},\, \hat{RH}_t,\, 
\hat u_t,\, \hat v_t)' ,
\]
where the elements of the vector are defined as before. Note that the difference from $\hat{\mathbf{x}}_t$ is the plug-in estimate for potential virtual temperature take the place of the variables over which is defined. The 
observation-uncertainty vector $\sigma_{\bar{\mathbf{x}},t}$ contains, the uncertainties provided by GRUAN, except for 
$\sigma_{\hat{\theta}_{v},t}$, which is obtained through the law of propagation of 
uncertainty.

The state vector is denoted by
\[
\mathbf{s}_t=(z_{t},\Lambda_{z,t},\theta_{v,t},\Lambda_{\theta_v,t},RH_{t},\Lambda_{RH,t},u_t,\Lambda_{u,t},v_t,\Lambda_{v,t})'
\]
where $\Lambda$, subscripted by a specific variable, denotes its rate of change. For example, $\Lambda_{z,t}=\partial z_t/\partial t$ is the radiosonde vertical ascent rate.

The observation equation specifies
\[
\begin{cases}
    \hat z_t=z_{t}+\epsilon_{z,t}\\
    \hat{\theta}_{v,t}=\theta_{v,t}+\epsilon_{\theta_v,t}\\
    \hat{RH}_t=RH_{t}+\epsilon_{RH,t}\\
    \hat u_t=u_t+\epsilon_{u,t}\\
    \hat v_t=v_t+\epsilon_{v,t}.\\
\end{cases}
\]
Only a subset of the state components is directly measured. The observable variables  are
$(z_t,\theta_{v,t},RH_t,u_t,v_t)$, while the remaining state elements 
$(\Lambda_{z,t},\Lambda_{\theta_v,t},\Lambda_{RH,t},\Lambda_{u,t},\Lambda_{v,t})$ 
are latent. 
The measurement disturbances $\boldsymbol{\epsilon}_t=(\epsilon_{z,t},\,\epsilon_{\theta_v,t},\,\epsilon_{RH,t},\,\epsilon_{u,t},\epsilon_{v,t})'$ are assumed independent, 
normally distributed with zero mean and diagonal covariance matrix $H_t$, representing the measurement uncertainty affecting observations, that is 
\[
\operatorname{diag}(H_t)
=(\sigma_{\hat{z},t}^2,\,\sigma_{\hat{T},t}^2,\,\sigma_{\hat{p},t}^2,\,\sigma_{\hat{RH},t}^2,\,\sigma_{\hat{r},t}^2,\,\sigma_{\hat{u},t}^2,\,\sigma_{\hat{v},t}^2)'.
\]
 A future possible extension might include correlated errors by assuming $H_t$ not diagonal.

The state equation takes the following form 
\[\begin{cases}
    z_t = z_{t-1} + \Lambda_{z,t-1} + \eta_{z,t},\\
    \Lambda_{z,t} = \Lambda_{z,t-1} + \eta_{\Lambda,z,t}\\
    \theta_{v,t} = \theta_{v,t-1} + \Lambda_{\theta_v,t-1} + \eta_{\theta_v,t}\\
    \Lambda_{\theta_v,t} = \Lambda_{\theta_v,t-1} + \eta_{\Lambda,\theta_v,t}\\
    RH_t = RH_{t-1} + \Lambda_{RH,t-1} + \eta_{RH,t}\\
    \Lambda_{RH,t} = \Lambda_{RH,t-1} + \eta_{\Lambda,RH,t},\\
    u_t = u_{t-1} + \Lambda_{u,t} + \eta_{u,t}\\
    \Lambda_{u,t} = \Lambda_{u,t-1} + \eta_{\Lambda,u,t}\\
    v_t = v_{t-1} + \Lambda_{v,t} + \eta_{v,t}\\
    \Lambda_{v,t} = \Lambda_{v,t-1} + \eta_{\Lambda,v,t}
\end{cases}\]
where $\mathbf{\eta}_t=(\eta_{z,t},\,\eta_{\Lambda,z,t},\,\eta_{\theta_v,t},\,\eta_{\Lambda,\theta_v,t},\,\eta_{RH,t},\,\eta_{\Lambda,RH,t},\,\eta_{u,t},\,\eta_{\Lambda,u,t} ,\,\eta_{v,t},\,\eta_{\Lambda,v,t})'$ is known as the state disturbance vector and it is assumed indipendent normally distributed with zero mean and diagonal covariance matrix $Q_t$, such that
\[
\operatorname{diag}(Q_t)=
(
\sigma^2_{\eta_{z,t}},\,
\sigma^2_{\eta_{\Lambda,z,t}},\,
\sigma^2_{\eta_{\theta_v,t}},\,
\sigma^2_{\eta_{\Lambda,\theta_v,t}},\,
\sigma^2_{\eta_{RH,t}},\,
\sigma^2_{\eta_{\Lambda,RH,t}},\,
\sigma^2_{\eta_{u,t}},\,
\sigma^2_{\eta_{\Lambda,u,t}},\,
\sigma^2_{\eta_{v,t}},\,
\sigma^2_{\eta_{\Lambda,v,t}}
)'.
\]
Such a model assumes a random walk evolution for all rates of change $\Lambda_.$, implying that each underlying variable follows a local–linear–trend (LLT) dynamics\cite{shumstof2000}. This means that both the level and its time derivative are allowed to vary stochastically through the state disturbances $\boldsymbol{\eta}_t$, providing a flexible structure capable of capturing smooth temporal variability as well as abrupt changes driven by the variance components in $Q_t$. 

The choice of this model, instead of a simpler local--linear--level (LLL) model \cite{shumstof2000}, in which derivatives are not included in the state equation and the states themselves are modeled as random walks, has the advantage of explicitly modeling the time derivatives and their uncertainty, which can then be exploited to estimate the gradients. Moreover, these smoothed results do not compromise the ability to estimate gradients, but rather suggest a robust gradient estimation under noise-affected observations. Analogously as before we denote by $\bar{X}_n=\{\bar{\mathbf{x}}_t:t=1,\dots,n\}$ and $S_n=\{\mathbf{s}_t:t=1,\dots,n\}$.

One of the main advantages of such a model is that, thanks to the well–established 
and widely applied Kalman smoothing technique, it is possible to estimate the 
conditional distribution of the unobserved states given the full set of 
observations. Formally, the Kalman smoother provides
\[
\bar{\mathbf{s}}_t = E[\mathbf{s}_t \mid \bar{X}_n]
    = (\bar{z}_{t},\bar{\Lambda}_{z,t},\bar{\theta}_{v,t},\bar{\Lambda}_{\theta_v,t}, \bar{RH}_{t},\bar{\Lambda}_{RH,t},\bar{u}_t,\bar{\Lambda}_{u,t}, \bar{v}_t,\bar{\Lambda}_{v,t})',
\]
known as the smoothed state vector, together with
\[
\bar{P}_t = E\!\left[(\mathbf{s}_t-\bar{\mathbf{s}}_t)
                 (\mathbf{s}_t-\bar{\mathbf{s}}_t)'\mid \bar{X}_n\right],
\]
the smoothed state covariance matrix. These quantities represent the 
posterior mean and posterior uncertainty of the latent state at time $t$ after the entire profile $\bar{X}_n$ has been assimilated.

Such a technique requires all SSM matrices to be known. In the present application, the only element not specified a priori is the disturbance covariance matrix $Q_t$. These quantities are estimated by maximizing the likelihood of the state-space model. The estimation proceeds iteratively: starting from an initial guess of the parameters, 
the Kalman filter and smoother are applied, the log-likelihood is evaluated, and the 
parameters are updated until convergence.

The initial parameter values are chosen to ensure numerical stability of the filter 
and to provide a reasonable starting point for the likelihood optimization. Although, these initial values are not critical—since the maximum-likelihood procedure rapidly adjusts them during the first iterations—they help avoid degeneracies in the early 
filtering steps.

Several numerical optimizers were tested for the maximization step, including
Nelder-Mead, limited-memory BFGS (LBFGS), and the modified Powell method. Across all profiles, the Powell algorithm consistently provided the most reliable performance, converging rapidly and yielding the highest log-likelihood values. LBFGS converged only intermittently and required substantially more iterations, while Nelder-Mead failed to converge in many cases. For this reason, the Powell optimizer was adopted.

Having obtained the smoothed state vectors after parameter estimation, it is 
possible to define the SSM–enriched diagnostic–variable vector estimates. 
Formally,
\begin{equation}
\bar{\mathbf{y}}_t = \bar{f}(\bar{\mathbf{s}}_t)
    = \bigl(\bar{\theta}_{v,t},\,
            \bar{\Gamma}_{\theta_v,t},\,
            \bar{\Gamma}_{RH,t},\,
            \bar{Ri}_t \bigr)',
\label{eq:smooth_y}
\end{equation}

where the vertical gradients are computed directly from the smoothed rates of 
change,
\[
\bar{\Gamma}_{\theta_v,t}
    = \frac{\bar{\Lambda}_{\theta_v,t}}{\bar{\Lambda}_{z,t}},
\qquad
\bar{\Gamma}_{RH,t}
    = \frac{\bar{\Lambda}_{RH,t}}{\bar{\Lambda}_{z,t}},
\]
and the Richardson number estimate is given by
\[
\bar{Ri}_t
    = \frac{(g/\bar{\theta}_{v,t})\,
            (\bar{\Lambda}_{\theta_v,t}/\bar{\Lambda}_{z,t})}
           {\left(\bar{\Lambda}_{u,t}/\bar{\Lambda}_{z,t}\right)^2
           +\left(\bar{\Lambda}_{v,t}/\bar{\Lambda}_{z,t}\right)^2}.
\]

Note that, because the function $\bar{f}$ in equation \ref{eq:smooth_y} is nonlinear, these quantities are not
the conditional expectations of the diagnostic variables themselves, but rather
plug–in estimators obtained by evaluating diagnostic functions in the
smoothed state $E[\mathbf{s}_t\mid \bar{X}_n]$.

The diagnostic variable uncertainty, $\sigma_{\bar{\mathbf{y}},t}$ can be estimated by applying the law of 
propagation of uncertainty to the smoothed state covariance $\bar{P}_t$, where the 
uncertainty of each smoothed state component is given by 
$\sigma_{\bar{\mathbf{s}},t}=\sqrt{\operatorname{diag}(\bar{P}_t)}$.

Finally, the smoothed estimate of the PBLH can be obtained from the 
diagnostic–variable estimates $\bar{Y}_n=\{\bar{\mathbf{y}}_t:t=1,\dots,n\}$ by applying the 
corresponding criterion function $g$ to the full set of smoothed diagnostics,
\begin{equation}
\bar{\mathbf{h}}
    = g\bigl(\bar{Y}_n\bigr)
    = (\bar h_{PM},\, 
       \bar h_{\theta_v},\, 
       \bar h_{RH},\, 
       \bar h_{RM})',
\label{eq:smooth_h}
\end{equation}

As stated before, regarding the standard plug-in estimation in equation~\ref{eq:plug-in_estimate}, 
since the derivatives required for the function $g$, which applies the PBLH criterion, cannot be 
computed analytically, the uncertainty of the PBLH estimate vector $\sigma_{\bar{\mathbf{h}}}$ 
cannot be directly inferred from $\{\sigma_{\bar{\mathbf{y}},t}:t=1,\dots,n\}$.

The procedure to estimate the full posterior distribution $p(\mathbf{h}\mid \bar{X}_n)$ is built on the 
state–space framework introduced above, and is detailed in the following section (\ref{sec:3.4}).

\subsection{Monte Carlo Simulation Estimation}
\label{sec:3.4}
As stated in \cite{JCGM2008}, when the GUM framework, i.e. the law of propagation of 
uncertainty, is not applicable, a common and robust alternative is the Monte Carlo (MC) 
approach. The overall procedure consists of generating $M$ synthetic atmospheric vertical 
profiles from their prescribed probability distributions. For each generated profile, a 
PBLH estimate is computed. The resulting ensemble of $M$ PBLH estimates provides an 
approximation of the PBLH distribution, namely the posterior distribution conditioned on 
the available observations.

Preliminary analyses suggest that the resulting distribution may be skewed or even 
multimodal. For this reason, we adopt the median as the point estimate and the 95\% 
percentile range as the expanded measure of uncertainty.

An advantage of the state-space model approach is that it also provides a solution for generating synthetic profiles thanks to simulation smoothing. In its simplest form, simulation smoothing consists of generating samples of the state vectors profile from its conditional distribution given the full set of observations, that is to draw
\[
\tilde{S}_n \sim p({S}_n\mid \bar{X}_n),
\]
so that each draw $\tilde{S}_n$ represents one possible realization of the states profile consistent with the observations and measurement uncertainties.

For this task, several backward–recursive algorithms are available to generate state simulations, most notably the classical Forward Filtering, Backward Sampling (FFBS) method and its more advanced variants, such as the simulation smoother introduced in \cite{simulation_smoother}. These algorithms produce 
samples from the conditional state distribution $p({S}_n\mid \bar{X}_n)$ by combining the Kalman filter with a backward sampling step that enforces dynamical consistency across the entire state trajectory.

Exploiting this capability, the method for MC PBLH estimation and quantification of its uncertainty proposed in this paper can be summarized by the following pseudo–algorithm.

\medskip
\noindent\textbf{PBLH MC estimate and uncertainty}
\label{alg:pblh_unc}

\noindent\textbf{Input:} observations $\bar{X}_n$ and measurement uncertainties 
$\{\sigma_{\bar{\mathbf{x}},t}: t=1,\dots,n\}$.

\begin{enumerate}
\item Estimate the unknown state–space model parameters $Q_t$ by maximum likelihood.
\item Generate $M$ independent simulations of the unobserved state vertical profile 
      using simulation smoothing,
\[
      \tilde{S}_n^{1},\dots,\tilde{S}_n^{M} \stackrel{\text{i.i.d.}}{\sim} p(S_n \mid \bar{X}_n).
      \]
\item For each $i=1,\dots,M$, compute the corresponding diagnostic–variable profile
\[
      \tilde{Y}_n^{i} = \{\,\bar{f}(\tilde{\mathbf{s}}_t^{i}) : t=1,\dots,n\,\},
      \]
      by applying equation~\ref{eq:smooth_y} to the simulated states.
\item For each $i=1,\dots,M$, compute the PBLH estimate by applying 
      equation~\ref{eq:smooth_h} to the simulated diagnostic variables,
\[
      \tilde{\mathbf{h}}^{i} = g(\tilde{Y}_n^{i}).
      \]
\item From the empirical distribution of the $M$ PBLH samples 
      \(\tilde{H}_M=\{\tilde{\mathbf{h}}^{i} : i=1,\dots,M\}\), 
      estimate the Monte Carlo PBLH as the median,
\[
      \mathbf{h}^{MC} = \operatorname{median}\{\tilde{H}_M\},
      \]
      and quantify its expanded uncertainty \(u(\mathbf{h}^{MC})\) using the 
      95\% percentile range.
\end{enumerate}
\noindent\textbf{Output:} MC PBLH estimate $\mathbf{h}^{MC}$ and its expanded uncertainty $u(\mathbf{h}^{MC})$.
\medskip

The number of simulations $M$ must be chosen to ensure a sufficiently accurate 
approximation while avoiding unnecessary computational cost. Empirical analyses 
indicate that $M = 200$ provides a reasonable trade-off between accuracy and 
computational efficiency, and this value is therefore adopted.

\begin{figure}[h!]
    \centering
    \includegraphics[width=\linewidth]{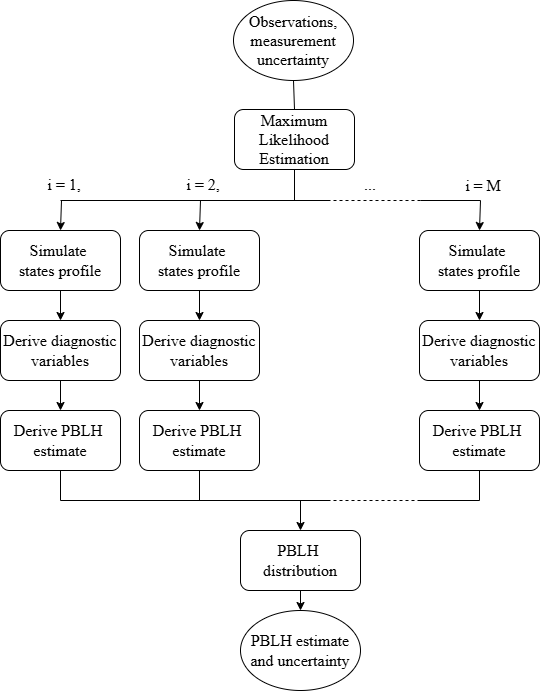}
    \caption{Flow chart of the PBLH Monte Carlo procedure.}
    \label{fig:flow_chart}
\end{figure}

Figure \ref{fig:flow_chart} shows the flow chart of the procedure.

\section{Case studies}
\label{sec:4}

Given the proposed methodology, we present a preliminary 
analysis of the dataset specified in Section~\ref{sec:2}. 

\subsection{Example profiles}
\label{sec:4.1}

A set of representative atmospheric profiles is examined in detail. These examples serve to clarify the methodological workflow and highlight the key diagnostic features used throughout the study.

\begin{figure}[!h]
    \centering
    \includegraphics[width=1\linewidth]{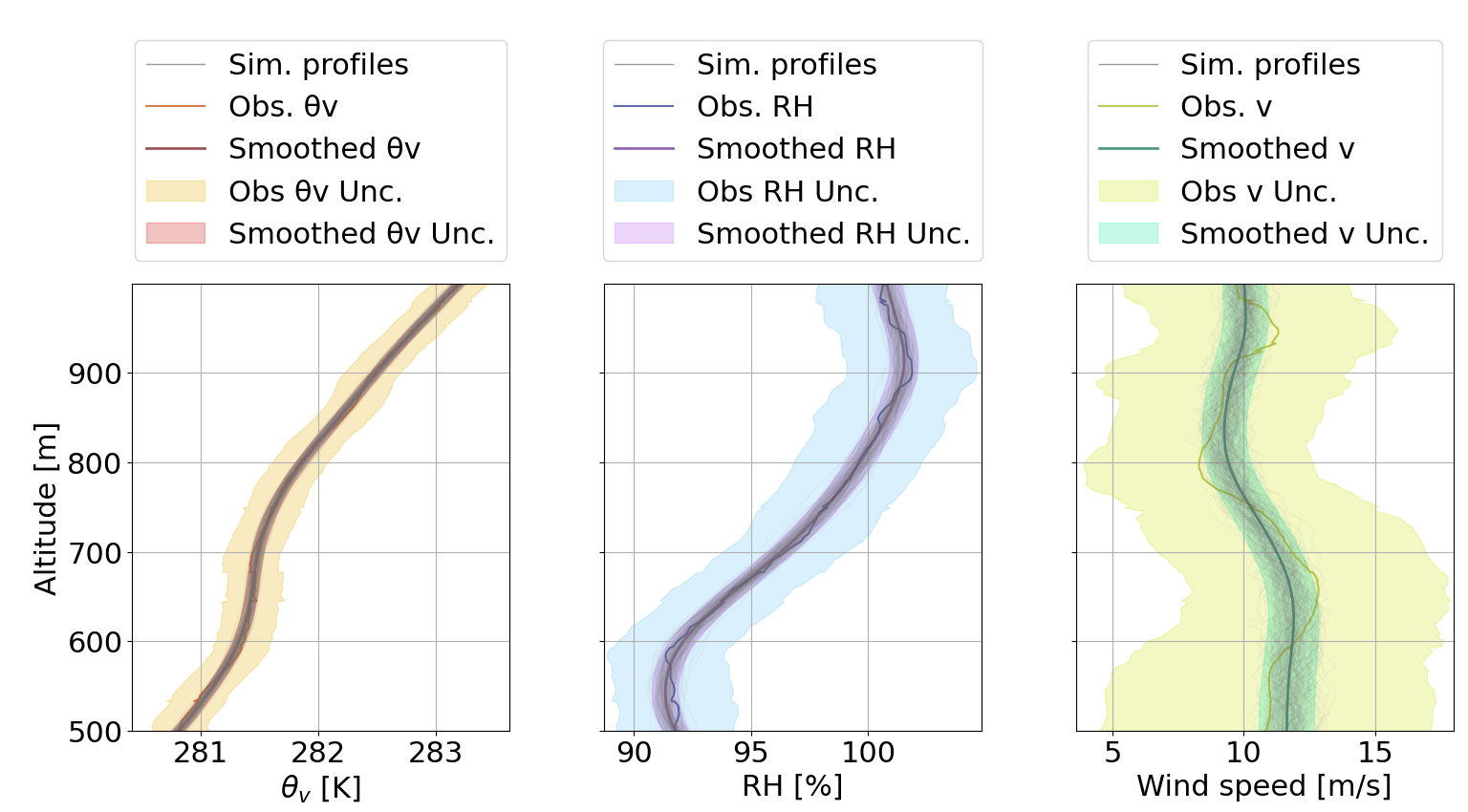}
    \caption{Example profile from the Lindenberg site. Comparison between observational measurement uncertainty and the 
    uncertainty of the smoothed state variables obtained from the SSM 
    framework. The shaded bands represent the expanded uncertainty intervals for 
    both the observations and the smoothed states. The smoothed uncertainties are 
    systematically smaller, reflecting the regularizing effect of the MLE-based 
    smoothing procedure and its ability to exploit the known measurement error 
    structure. Simulated state trajectories are also shown and remain well contained within the 
    smoothed-state uncertainty envelopes.}
    \label{fig:unc_zoom}
\end{figure}

Figure \ref{fig:unc_zoom} shows an interesting first result: the uncertainty after 
applying the Kalman smoother is always smaller than the original measurement error. 
This means that the smoother does not simply clean up the GRUAN vertical profile; 
it also exploits the known sensor error characteristics to reconstruct a realistic 
atmospheric trajectory while reducing noise. As expected, the simulated profiles 
remain contained within these narrower uncertainty envelopes, confirming that the 
algorithm behaves consistently with the underlying state-space model assumptions.

These results were obtained using a modified Powell's method, implemented through 
the `scipy.optimize` solver and wrapped by the Statsmodels API for maximum 
likelihood estimation. The optimization converged successfully after three 
iterations and 418 function evaluations. Further analysis will include a detailed assessment 
of the residual distribution as well as checks for potential oversmoothing in the state-space reconstruction.

To illustrate the applied methodology, we consider three representative atmospheric profiles, two from each station for daytime and nighttime, respectively.

\begin{figure}[!h]
    \centering
    \includegraphics[width=\linewidth]{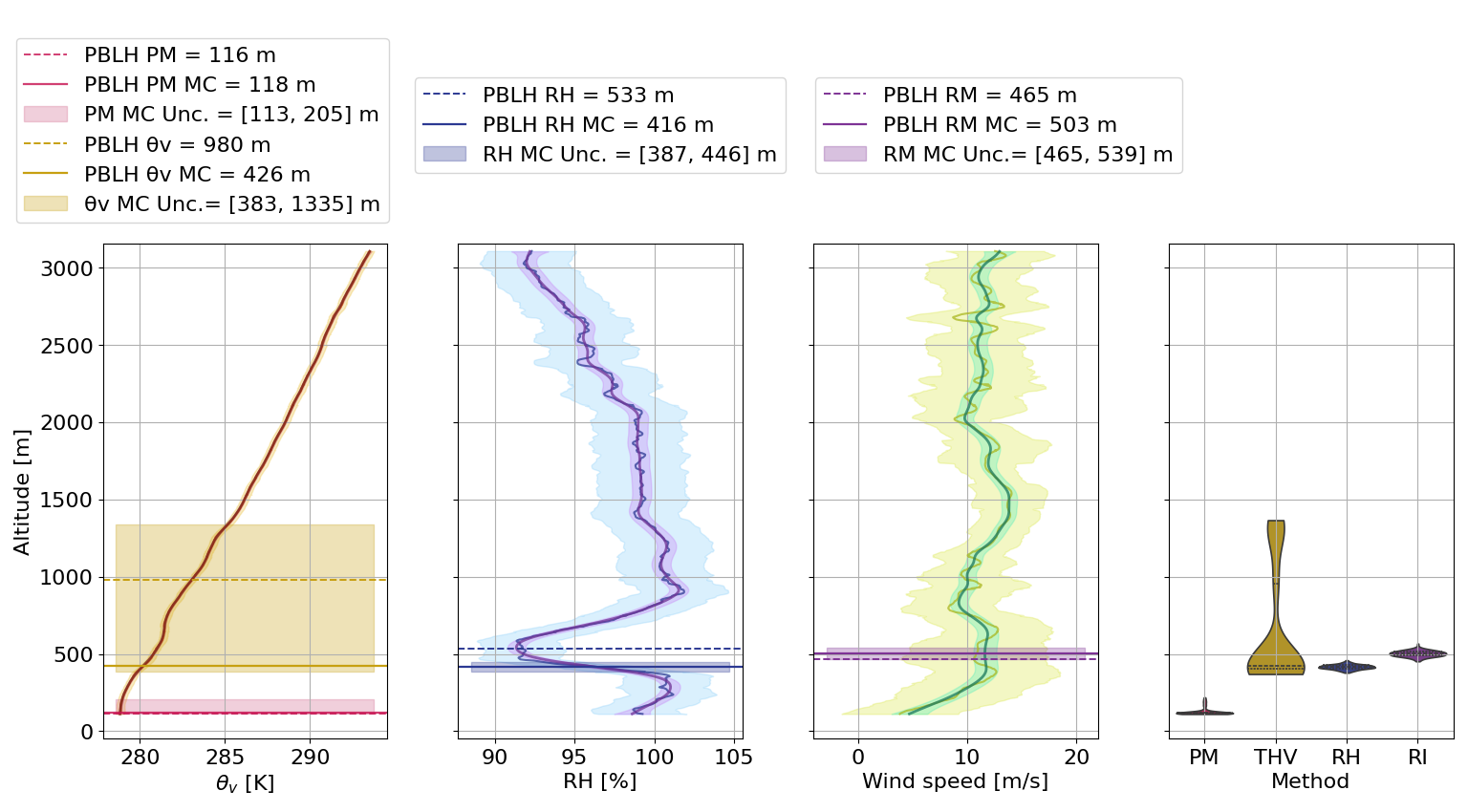}
    \caption{Profile from LIN during daytime (Local Launch Time: 2024-01-02 11:45:51 CET), including PBLH estimates from 
    standard plug-in methods and Monte-Carlo (MC) estimates and uncertainty. Standard 
    estimates are shown as dotted horizontal lines, while MC median estimates 
    appear as solid horizontal lines. MC-derived PBLH uncertainty is represented 
    by rectangles spanning the 95\% quantile range of the MC distribution.
    Left: observed and smoothed virtual potential temperature with 
    uncertainty bands, together with PBLH estimates from the PM and 
    $\theta_v$ methods and MC uncertainty ranges.
    Middle-left: relative humidity profile, RH-based PBLH estimates, and 
    associated uncertainty.
    Middle-right: meridional wind-speed component and PBLH estimates from the Richardson 
    number method.
    Right: violin plots of the MC PBLH distribution.
    Vertical profiles of $\theta_v$, $RH$ and meridional wind speed $v$ are presented as in Figure \ref{fig:unc_zoom}.}
    \label{fig:lin_ex_day}
\end{figure}

The Lindenberg daytime example in Figure~\ref{fig:lin_ex_day} exhibits several 
notable features. First, the standard plug-in estimates are generally consistent 
with the MC-based estimates, as they fall within the MC uncertainty range for all 
methods except the RH method, where the standard estimate lies slightly above the 
MC acceptance region. Three of the methods provide mutually coherent PBLH 
estimates, while the PM method identifies the PBLH at the surface due to the 
strictly increasing virtual potential temperature profile—highlighting that, 
under stable boundary-layer (SBL) conditions, the PM method is not well suited. 
The RH method successfully captures the major drop in relative humidity, and the 
meridional wind component also shows a change in behaviour above the PBLH 
identified by the Richardson-number method. It is also worth noting that the 
$\theta_v$ method exhibits relatively large uncertainty, driven by MC 
realizations that place the PBLH at higher altitudes, consistent with the absence 
of a single dominant jump in virtual potential temperature.

\begin{figure}[!h]
    \centering
    \includegraphics[width=\linewidth]{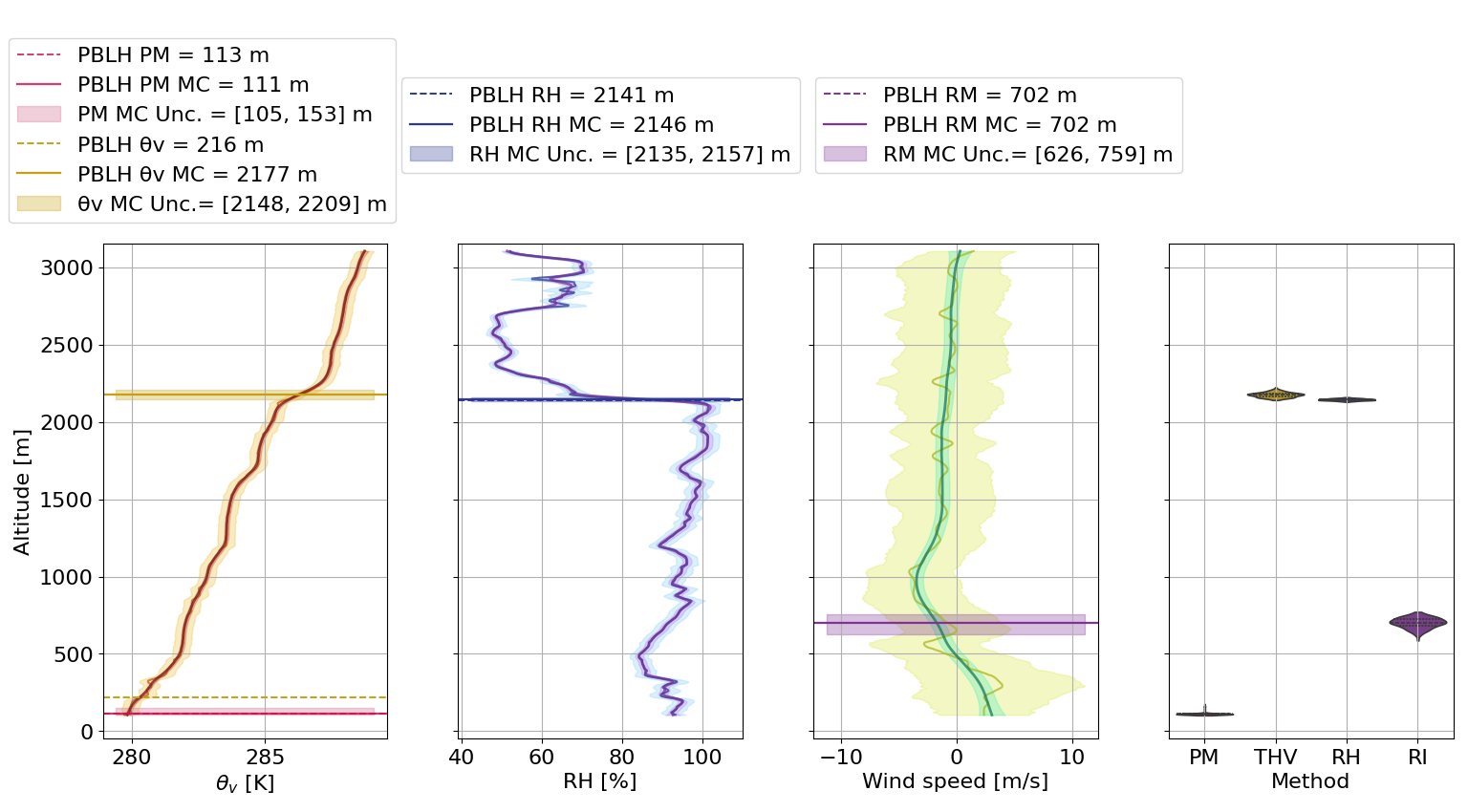}
    \caption{Profile from LIN during nighttime (Local Launch Time: 2024-01-01 23:45:10 CET), including PBLH estimates. For detailed descriptions, refer to the caption of 
    Figure~\ref{fig:lin_ex_day}.}
    \label{fig:lin_ex_night}
\end{figure}

The Lindenberg nighttime example in Figure~\ref{fig:lin_ex_night} illustrates a 
markedly different diagnostic situation. The PM method retrieves a very shallow 
nocturnal boundary layer (NBL), whereas the gradient-based methods detect a 
pronounced transition in both temperature and humidity, likely associated with 
the presence of a residual boundary layer (RBL) and the elevated capping 
inversion remaining from the previous day. The Richardson-number method again 
identifies an altitude where the wind field changes behaviour, consistent with a 
dynamical transition in the flow. Standard plug-in estimates remain broadly 
consistent with the MC-based estimates, and the MC uncertainty ranges are 
comparatively narrow across all methods. The only notable discrepancy arises for 
the $\theta_v$ method: the standard estimate is biased by high-resolution 
measurement noise in the turbulent lowest part of the atmosphere, which produces 
an artificially large gradient and leads to an erroneously low PBLH estimate.

The daytime Hong Kong example exhibits larger discrepancies between the standard 
plug-in estimates and the MC-based estimates. The $\theta_v$ method shows a 
particularly pronounced difference: the standard estimate fails to capture the 
strong increase in virtual potential temperature because the finite-difference 
gradient computed from high-resolution data is too noisy to resolve the true 
inversion structure. The RH method displays substantial uncertainty, reflecting 
the multiple humidity drops identified across the MC realizations. As this is 
the first example of a convective boundary layer (CBL), the PM estimate is 
especially informative. The Richardson-number method retrieves a PBLH consistent 
with the PM estimate and identifies a corresponding change in wind regime. 
Similar to the nighttime example in Figure~\ref{fig:lin_ex_night}, the 
gradient-based methods ($\theta_v$, RH) tend to detect a different transition 
level compared with the parcel and Richardson methods.

\begin{figure}[!h]
    \centering
    \includegraphics[width=1\linewidth]{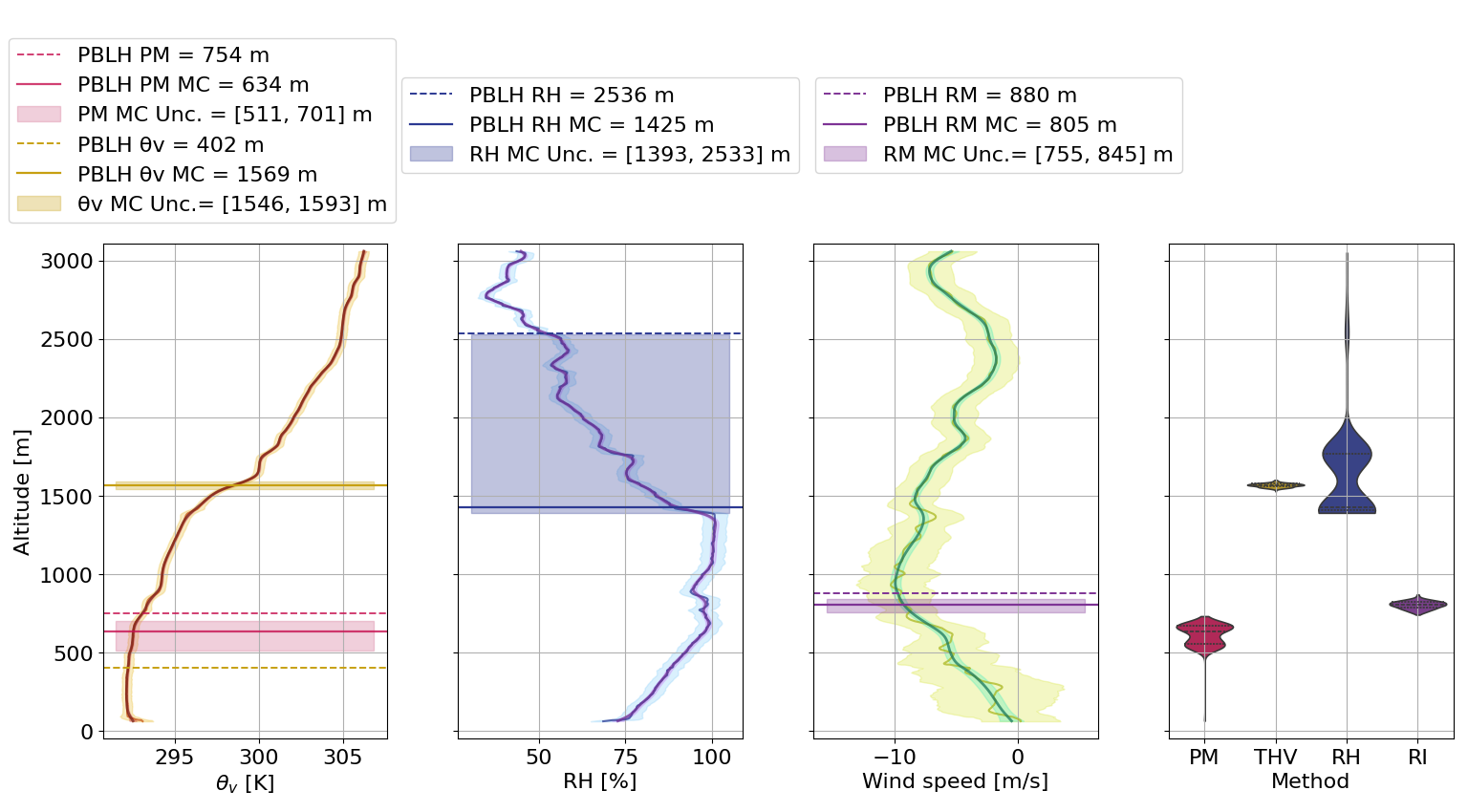}
    \caption{Profile from HKO during daytime (Local Launch Time: 2024-01-03 07:20:09 HKT), including PBLH estimates. For detailed descriptions, refer to the caption of 
    Figure~\ref{fig:lin_ex_day}.}
    \label{fig:hko_ex_day}
\end{figure}

The nighttime Hong Kong example shows strong agreement among the PM, 
$\theta_v$, and Richardson-number methods, while the RH method is again 
affected by the presence of multiple humidity drops. Reducing the maximum 
altitude considered in the RH search window would likely improve its 
performance by excluding spurious upper-level features. The temperature 
profile exhibits a clear stable nocturnal structure, and the main humidity 
transition is well defined, whereas changes in the wind regime are present 
but less pronounced. Both standard applications of the gradient-based 
methods result in very low PBLH estimates because the high-resolution 
finite-difference gradients are strongly perturbed by measurement noise in 
the turbulent lowest part of the atmosphere, producing artificially sharp 
gradients and consequently biased PBLH retrievals.

\begin{figure}[!h]
    \centering
    \includegraphics[width=1\linewidth]{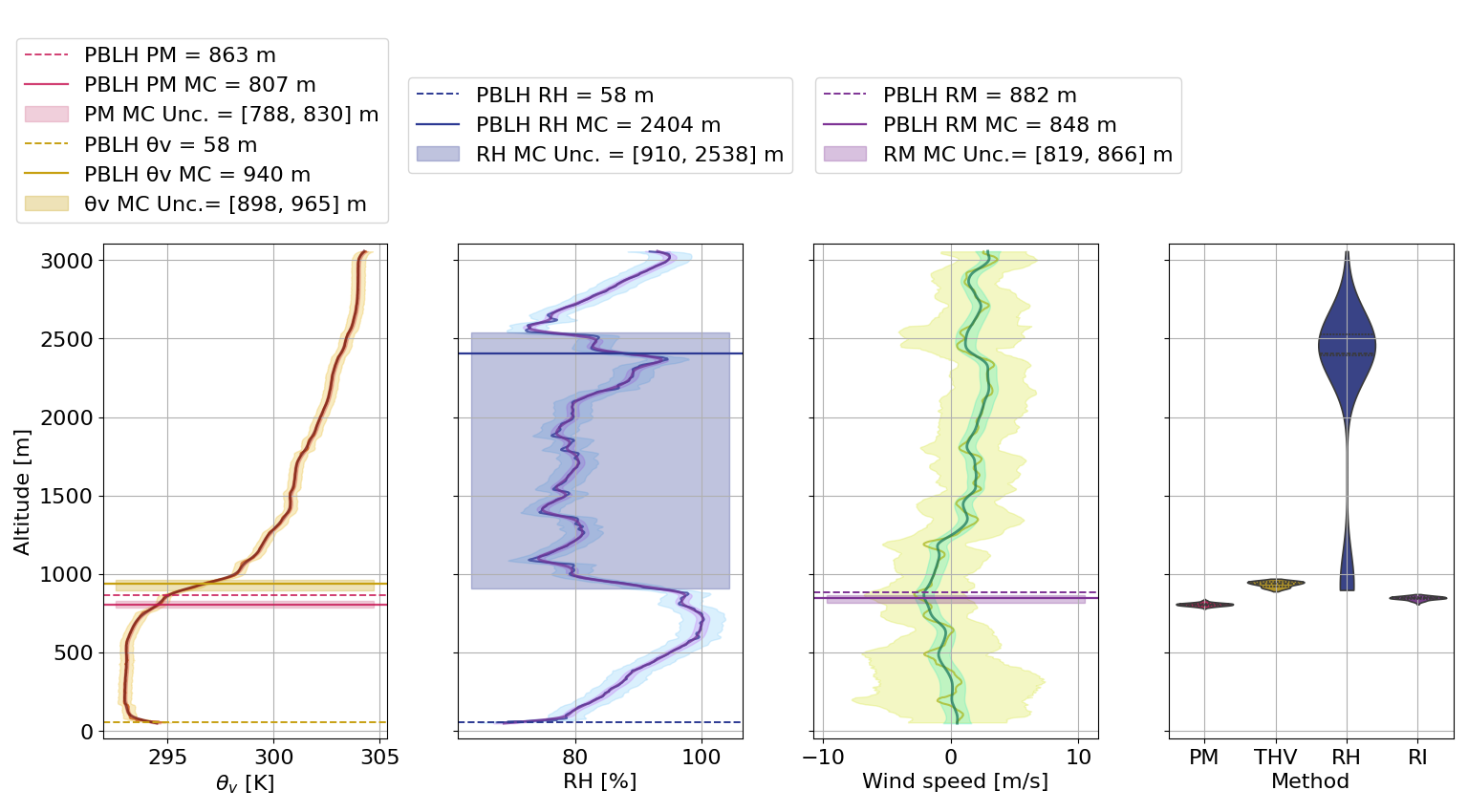}
    \caption{Profile from HKO during nighttime (Local Launch Time: 2024-01-01 19:20:05 HKT), including PBLH estimates. For detailed descriptions, refer to the caption of 
    Figure~\ref{fig:lin_ex_day}.}
    \label{fig:hko_ex_night}
\end{figure}

\begin{figure}[!h]
    \centering
    \includegraphics[width=1\linewidth]{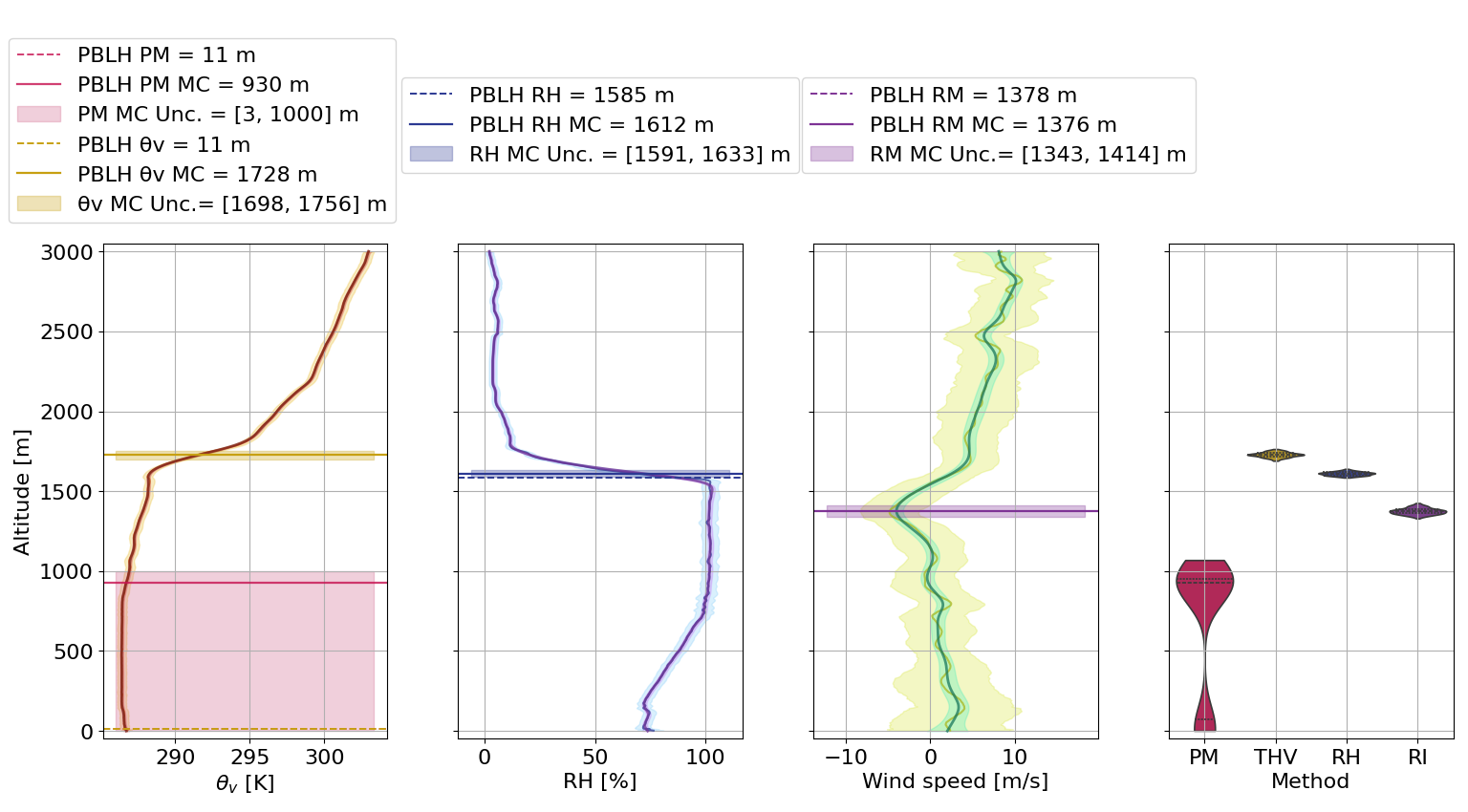}
    \caption{Profile from LAU during daytime (Local Launch Time: 2024-01-06 10:31:42 NZDT), including PBLH estimates. For detailed descriptions, refer to the caption of 
    Figure~\ref{fig:lin_ex_day}.}
    \label{fig:lau_ex_day}
\end{figure}

The daytime Lauder example in Figure~\ref{fig:lau_ex_day} shows strong 
consistency among the $\theta_v$, RH, and Richardson-number methods, both 
between the standard and MC approaches and across the methods themselves. All 
three diagnostics capture a clear transition in temperature, humidity, and wind, 
indicating a well-defined change of regime. Despite this overall coherence, the 
standard $\theta_v$ estimate is again misplaced, yielding an unrealistically low 
PBLH due to the perturbed finite-difference gradient derived from high-resolution 
data. The MC implementation of the PM method correctly identifies the mixed-layer 
height (MH), in contrast to the standard application, although its uncertainty 
range is broadened by a subset of MC realizations that place the PBLH near the 
surface. This behaviour mirrors the tendency of gradient-based methods to be 
biased downward when high-resolution turbulence contaminates the near-surface 
gradient estimates.

\begin{figure}[!h]
    \centering
    \includegraphics[width=1\linewidth]{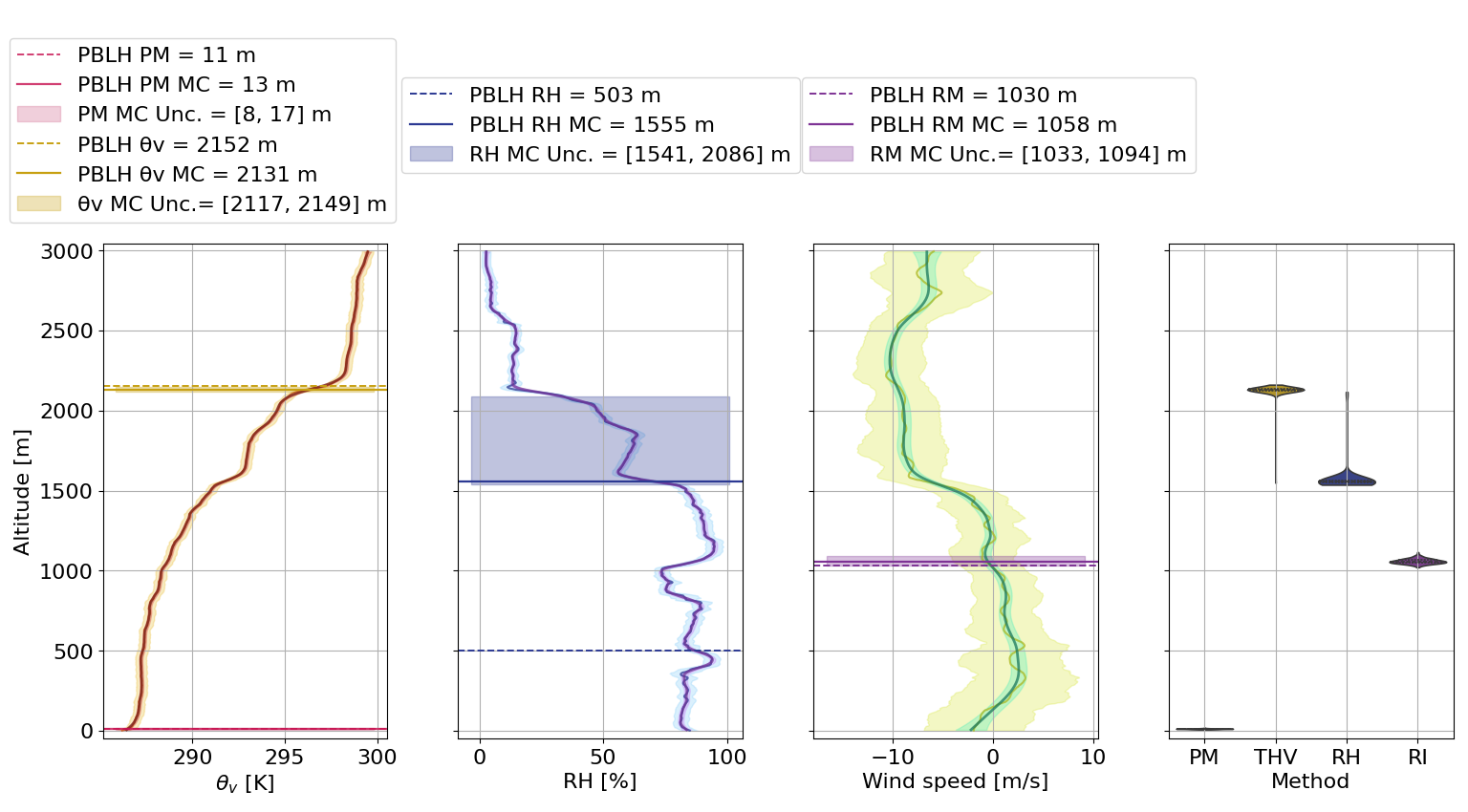}
    \caption{Profile from LAU during nighttime (Local Launch Time: 2024-01-04 22:31:04 NZDT), including PBLH estimates. For detailed descriptions, refer to the caption of 
    Figure~\ref{fig:lin_ex_day}.}
    \label{fig:lau_ex_night}
\end{figure}

The nighttime Lauder example in Figure~\ref{fig:lau_ex_night} exhibits consistent 
yet clearly distinct PBLH estimates across the different methods. The PM method 
detects the nocturnal boundary layer (NBL), while the gradient-based methods 
identify the lower and upper bounds of the previous day's capping inversion 
located above the residual layer. As in earlier examples, multiple humidity drops 
increase the uncertainty of the RH method and compromise the reliability of its 
standard plug-in estimate. The Richardson-number method retrieves a transition 
level associated with a slight change in wind behaviour and a sharp increase in 
humidity. However, a more pronounced overall change in all variables occurs at 
the altitude identified by the RH MC estimate, suggesting that the RH statistical 
diagnostic is capturing a physically meaningful transition that is less evident 
in the standard gradient-based approaches. 

\subsection{Aggregated Analysis}
\label{sec:4.2}

To infer broader properties of the proposed methodology relative to the 
standard plug-in estimation and its associated uncertainty, we compute 
summary statistics of the PBLH estimates and their uncertainties over the 
entire dataset. Results for each station are presented and discussed 
independently. Tables reporting the median and interquartile range (IQR) 
of all diagnostic quantities are provided in Section~\ref{sec:tbls}.

\begin{figure}[!h]
    \centering
    \hspace*{-1cm}
    \includegraphics[width=1.1\linewidth]{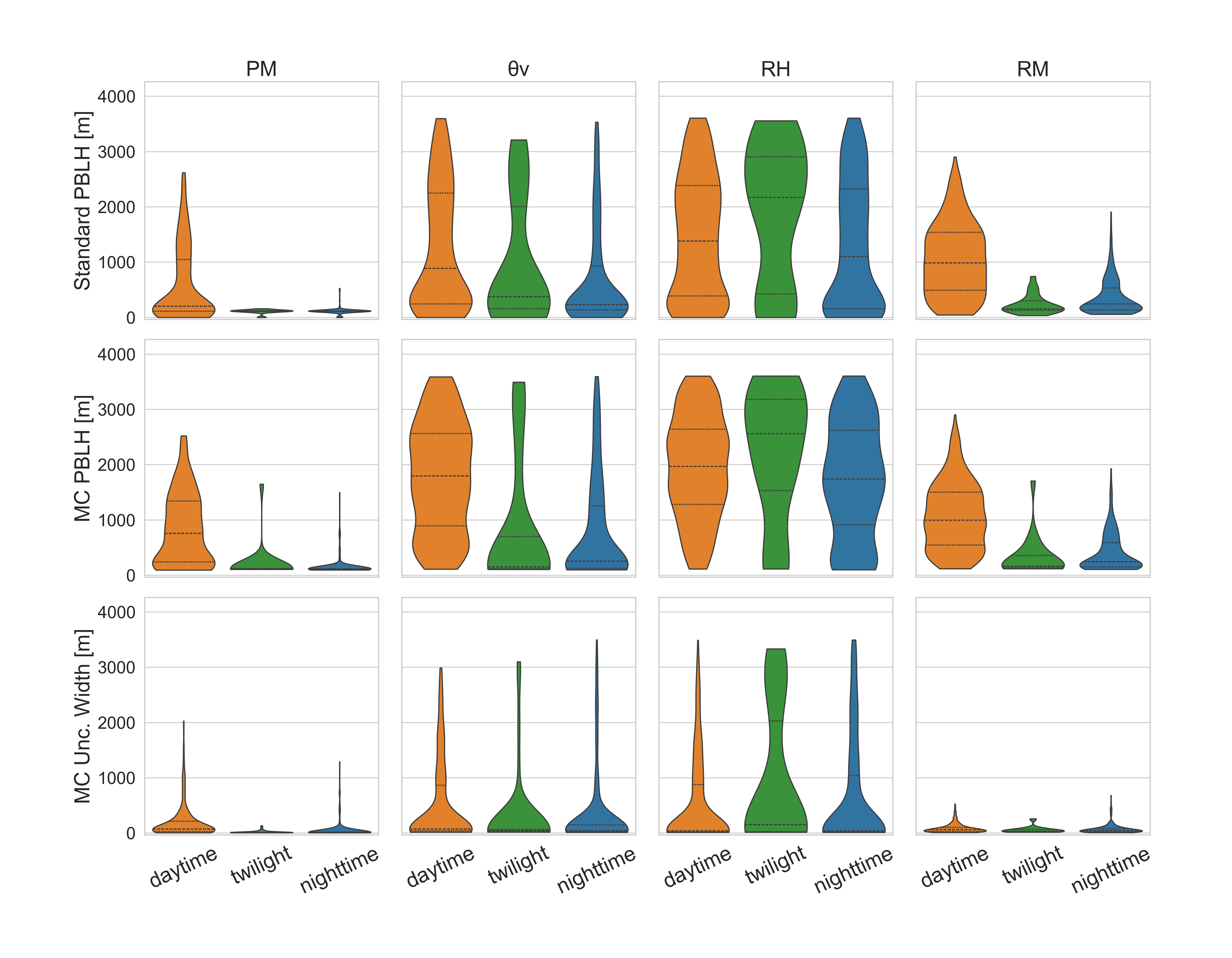}
    \caption{Violin plots for all Lindenberg profiles, grouped by time of day. 
    Each column corresponds to a different diagnostic method (PM, $\theta_v$, RH, 
    and RM), while each row corresponds to a different quantity: the standard 
    plug-in PBLH estimate, the MC PBLH estimate, and the MC 95\% 
    uncertainty width.}
    \label{fig:lin_violin_tod}
\end{figure}

We first aggregate the distributions with respect to the 
time of day (TOD). Starting with Lindenberg site, figures~\ref{fig:lin_violin_tod} and Table ~\ref{tbl:lin_tod} present the distributions of the standard plug-in estimates, the 
MC-based PBLH estimates, and the corresponding uncertainty widths for all 
diagnostic methods.
Different patterns are clearly recognizable. All methods exhibit the expected 
tendency for the PBLH to decrease from daytime to twilight and nighttime 
conditions. The different methods suggest progressively higher PBLH distributions in the 
following order: PM, RM, $\theta_v$, and RH.
Meanwhile, the MC estimates tend to increase and distribute the PBLH more 
evenly compared with the standard application. The only exception is the RM 
method, whose distribution remains particularly similar across the two 
approaches. This tendency is consistent with the behaviour observed in the previous 
examples, where the standard methods often produced misleading results. The 
uncertainty increases as the overall MC-estimated PBLH distribution rises, yet 
the MC approach still yields narrow and generally low uncertainty ranges. In 
contrast, some of the standard-method distributions have very low distribution 
but exhibit numerous outliers, particularly for the RH method during twilight, 
and to a lesser extent for the $\theta_v$ method.

\begin{figure}[!h]
    \centering
    \hspace*{-1cm}
    \includegraphics[width=1.1\linewidth]{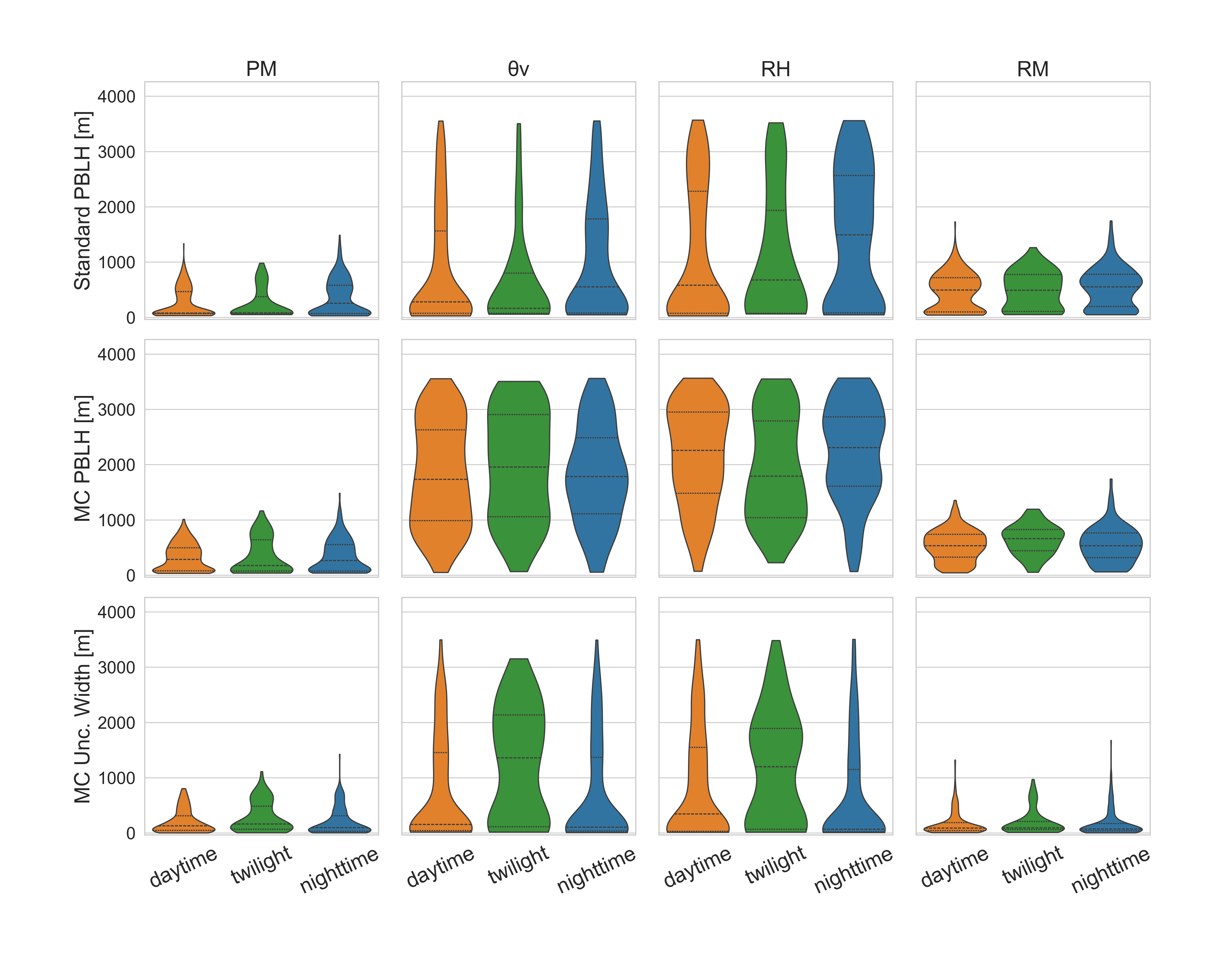}
    \caption{Violin plots for all Hong Kong profiles, grouped by time of day. For details, look at the caption of Figure ~\ref{fig:lin_violin_tod}.}
    \label{fig:hko_violin_tod}
\end{figure}

The Hong Kong site, shown in Figure~\ref{fig:hko_violin_tod} and 
Table~\ref{tbl:hko_tod}, displays both differences and points of consistency 
with the Lindenberg results. The daily cycle is much less pronounced, likely 
due to the tropical climate of the region, which reduces the contrast between 
daytime and nighttime boundary-layer regimes. The ordering of increasing PBLH estimates across methods remains the same, but 
in Hong Kong the gradient-based methods produce much higher PBLH values, whereas 
the PM and RM methods yield substantially lower distributions. The MC estimates again raise the barycentre of the distributions relative to the 
standard application—this effect is visible even for the RM method, but it is 
particularly pronounced for the gradient-based diagnostics. Aside from the PM 
method, the MC approach also appears to suppress much of the near-surface 
concentration present in the standard distributions, effectively removing a 
substantial portion of the low-level estimates.
Uncertainties appear contained with respect to the previous 
site, although the gradient-based methods exhibit a noticeably 
larger uncertainty width under twilight conditions.

\begin{figure}[!h]
    \centering
    \hspace*{-1cm}
    \includegraphics[width=1.1\linewidth]{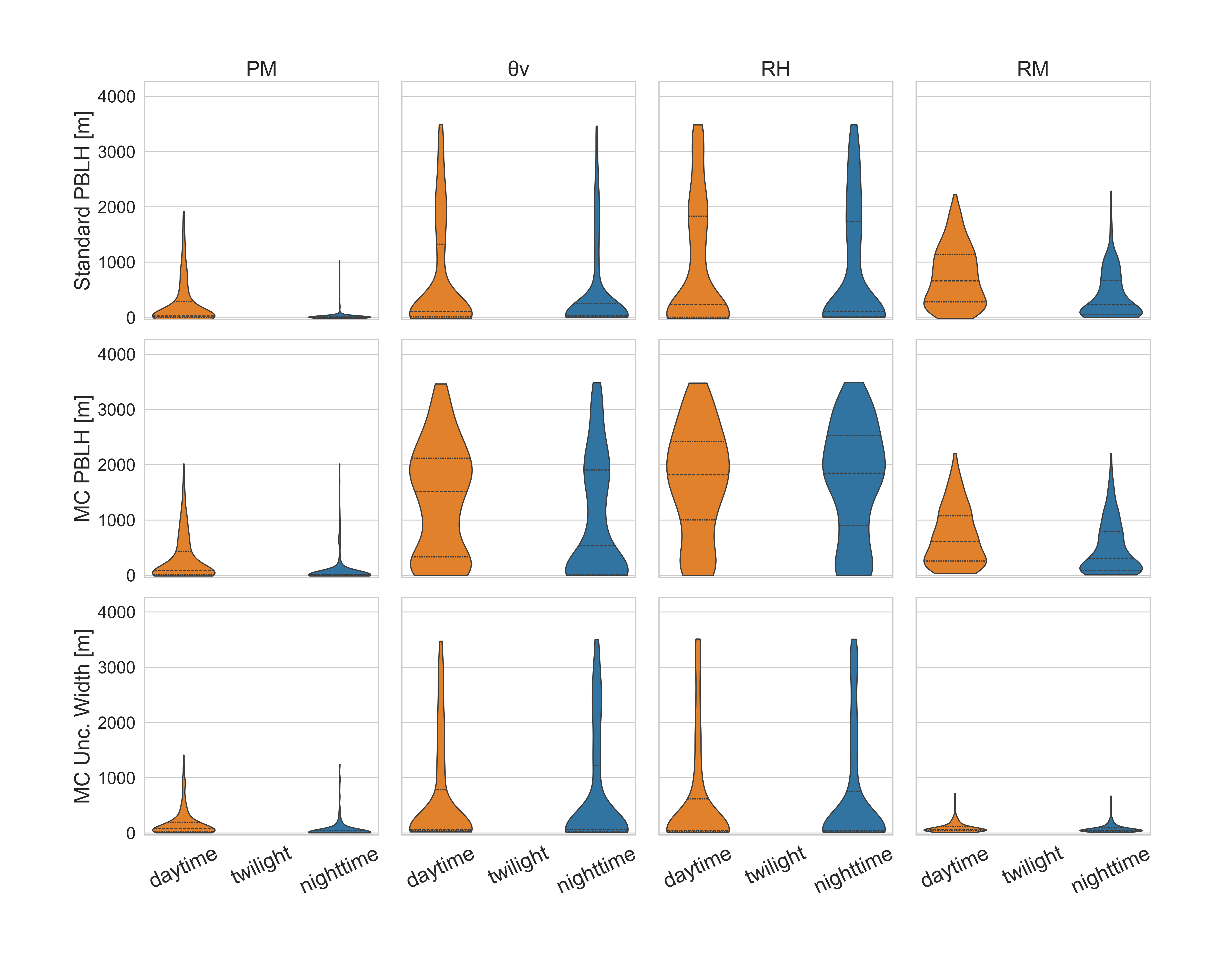}
    \caption{Violin plots for all Lauder profiles, grouped by time of day. There are no twilight profiles. For details, look at the caption of Figure ~\ref{fig:lin_violin_tod}.}
    \label{fig:lau_violin_tod}
\end{figure}

The Lauder site, shown in Figure~\ref{fig:lau_violin_tod} and 
Table~\ref{tbl:lau_tod}, exhibits behaviour broadly consistent with the 
Lindenberg results. Note that no profiles are available for twilight conditions. The 
daily cycle is clearly pronounced for most methods, with the exception of the 
RH diagnostic, which shows a much weaker diurnal modulation. The ordering of 
methods with respect to increasing PBLH estimates is consistent: PM, RM, 
$\theta_v$, and RH.

Using the MC approach instead of the standard plug-in application lifts the 
distributions, particularly for the gradient-based methods, which display both 
higher estimates and larger uncertainty widths. The RM uncertainty appears 
almost fixed across different times of day, showing limited sensitivity to the 
diurnal cycle.

\begin{figure}[!h]
    \centering
    \hspace*{-1cm}
    \includegraphics[width=1.1\linewidth]{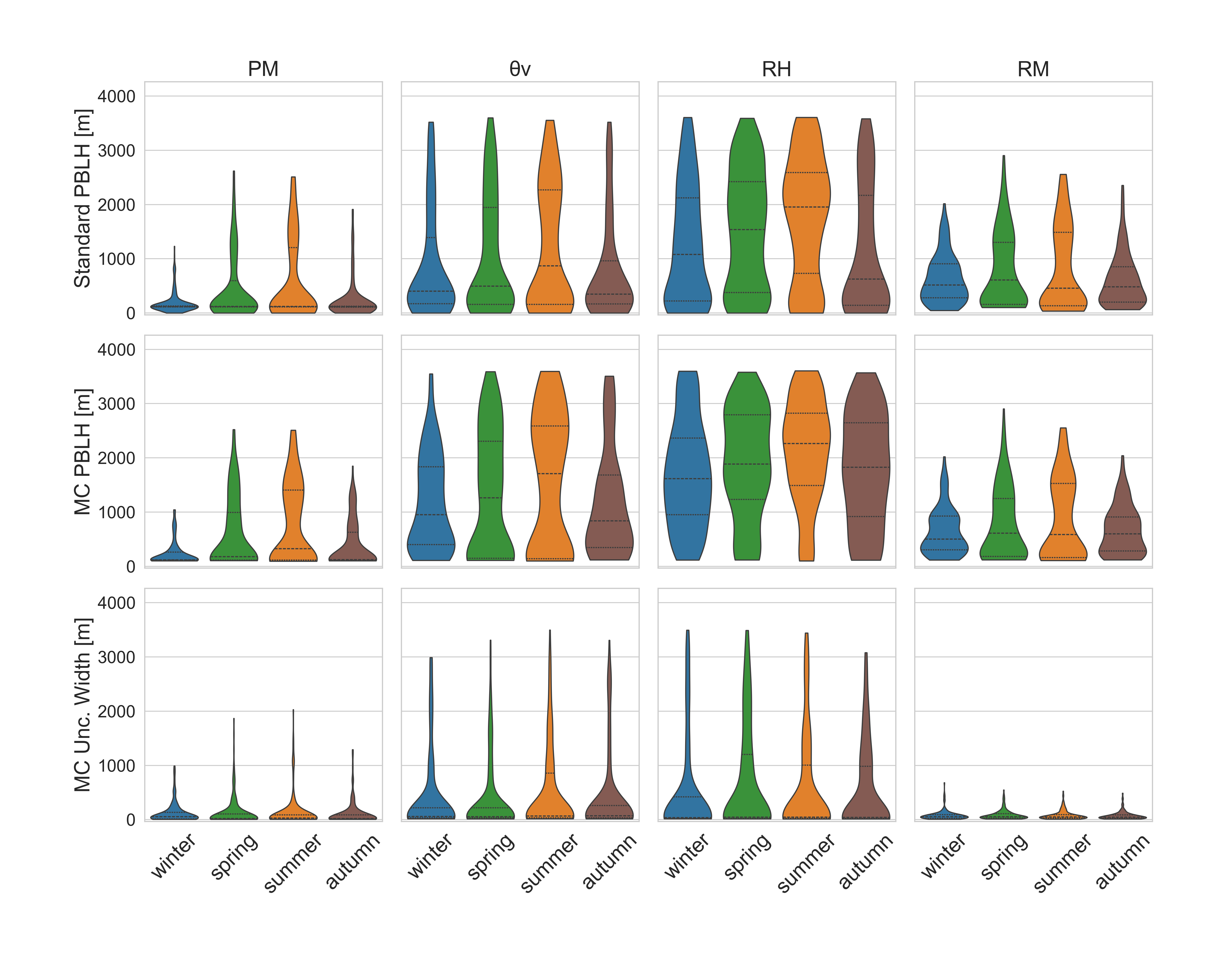}
    \caption{Violin plots for all Lindenberg profiles, grouped by season.
    Each column corresponds to a different diagnostic method (PM, $\theta_v$, RH, 
    and RM), while each row corresponds to a different quantity: the standard 
    plug-in PBLH estimate, the MC median PBLH estimate, and the MC 95\% 
    uncertainty width.}
    \label{fig:lin_violin_season}
\end{figure}

Grouping the profiles by season rather than by time of day again reveals 
interesting patterns. Figure~\ref{fig:lin_violin_season} and Table~\ref{tbl:lin_season_stats} show 
the seasonal distributions of the standard PBLH estimates, the MC-based 
estimates, and the MC uncertainty widths for the Lindenberg site. A clear annual cycle is evident across all methods, with minimum PBLH values 
occurring in winter and maximum values in summer.
The implications of using different PBLH diagnostics—or applying the MC 
approach instead of the standard plug-in method—remain consistent with the 
patterns discussed earlier.
The RM uncertainty appears largely independent of season and remains 
consistently narrow across all seasonal conditions.

\begin{figure}[!h]
    \centering
    \hspace*{-1cm}
    \includegraphics[width=1.1\linewidth]{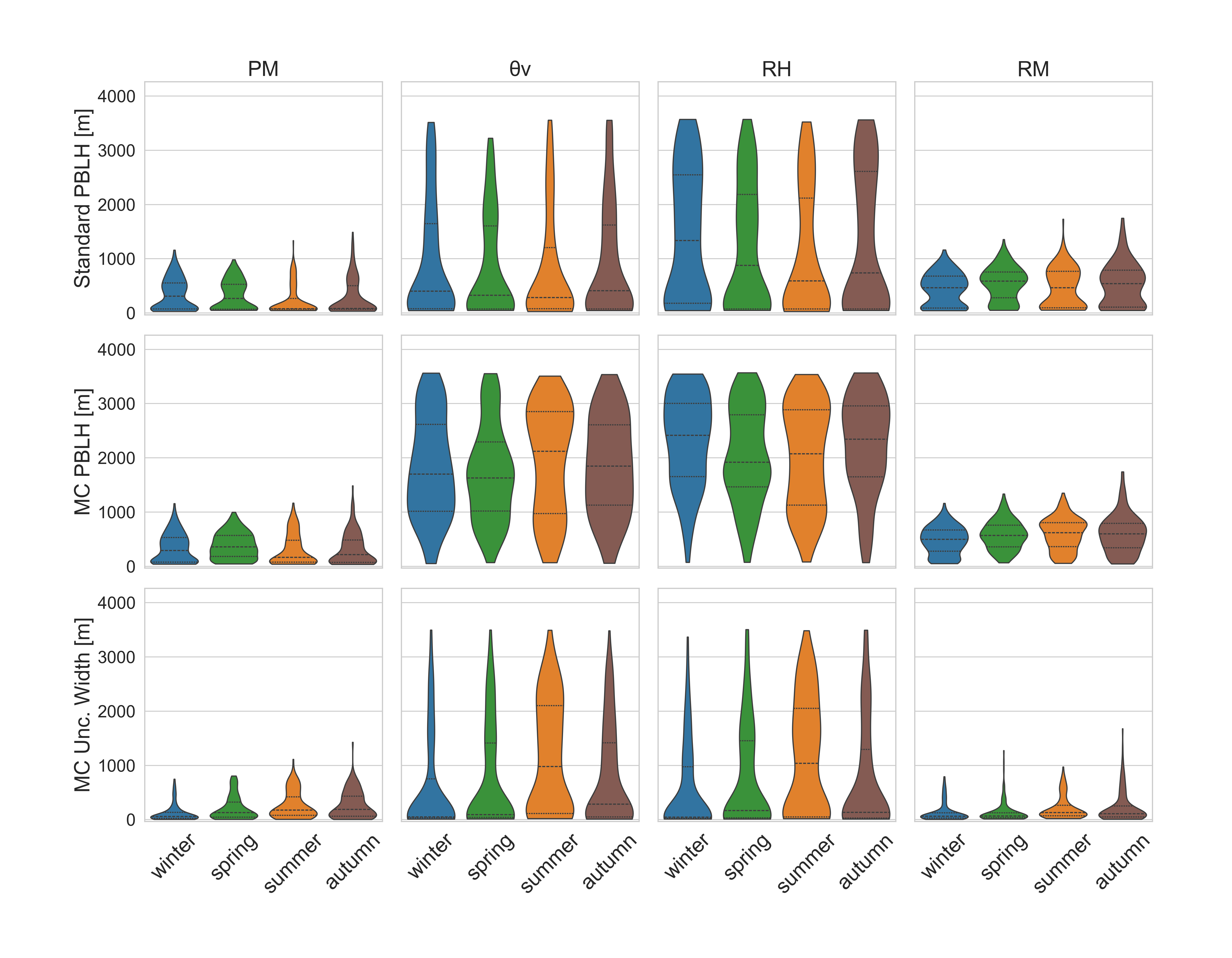}
    \caption{Violin plots for all Hong Kong profiles, grouped by season. For details, look at the caption of Figure ~\ref{fig:lin_violin_season}.}
    \label{fig:hko_violin_season}
\end{figure}

The seasonal cycle is much less evident in the Hong Kong case, as shown in 
Figure~\ref{fig:hko_violin_season} and Table~\ref{tbl:hko_season_stats}. The weak seasonality reflects the limited thermal contrast typical of 
subtropical--tropical climates, which reduces the separation between winter 
and summer boundary-layer regimes. The specific application of different PBLH 
diagnostics results in either distinct or relatively homogeneous seasonal 
distributions. Under the standard approach, winter and spring appear broadly 
consistent across methods; however, this behaviour does not hold when applying 
the MC framework. In particular, the MC versions of the PM and RM methods 
produce markedly different seasonal distributions, indicating that the 
probabilistic treatment amplifies seasonal contrasts that are muted in the 
standard plug-in estimates. Summer appears substantially more uncertain than the other seasons, particularly 
for the gradient-based methods, which exhibit the widest uncertainty widths 
during this period. Winter uncertainty is very narrow for both the PM and RM methods, indicating 
that these diagnostics remain highly constrained during the colder season.

\begin{figure}[!h]
    \centering
    \hspace*{-1cm}
    \includegraphics[width=1.1\linewidth]{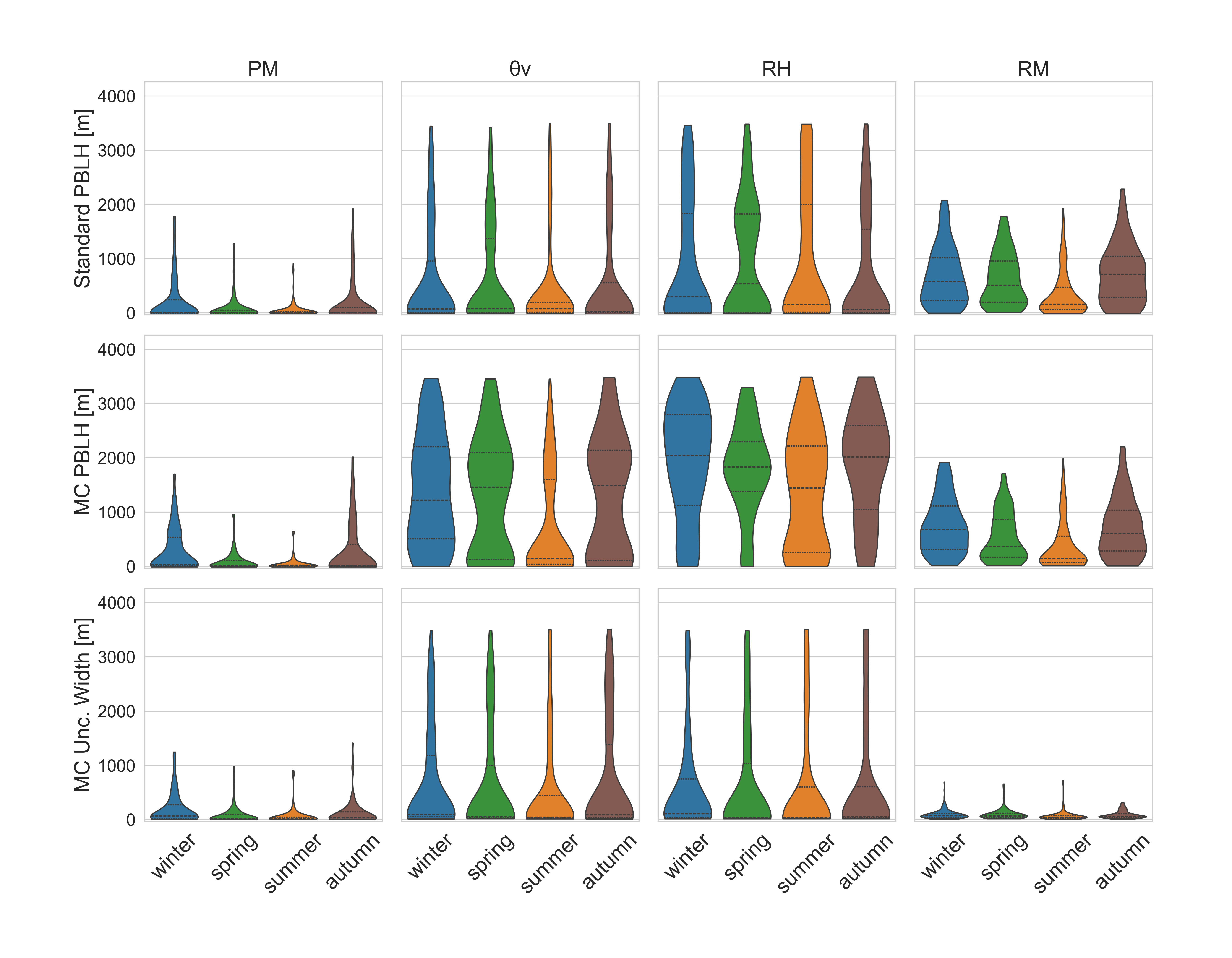}
    \caption{Violin plots for all Lauder profiles, grouped by season. For details, look at the caption of Figure ~\ref{fig:lin_violin_season}.}
    \label{fig:lau_violin_season}
\end{figure}

The Lauder site, shown in Figure~\ref{fig:lau_violin_season} and 
Table~\ref{tbl:lau_season_stats}, again appears broadly consistent with the 
Lindenberg example, though with some notable differences. The seasonal cycle is 
less pronounced, even if still clearly identifiable. Because the seasonal 
grouping was kept identical across hemispheres—defining winter as 
December--January--February—the maximum PBLH appears in “winter,” which 
corresponds to the local summer, while the minimum occurs in “summer,” 
corresponding to the local winter. This hemispheric inversion explains why the 
seasonal pattern differs from mid-latitude Northern Hemisphere sites. Overall, 
the Lauder distributions look more uniform throughout the year, with weaker 
contrasts between seasons.
The RM method appears largely agnostic with respect to season, showing only a 
slightly lower distribution during “winter” (i.e., the local summer). Apart 
from this modest dip, its behaviour remains remarkably stable throughout the 
year, reinforcing the method’s weak sensitivity to seasonal forcing at the 
Lauder site.
The contrast between the PM and RM methods and the gradient-based diagnostics 
is very pronounced. The gradient methods consistently produce higher PBLH 
estimates and substantially broader uncertainty distributions, highlighting 
their stronger sensitivity to atmospheric variability. In comparison, PM and RM 
remain much more constrained, with narrower and lower distributions across all 
seasons.

\section{Discussion and Future Development}
\label{sec:5}

The proposed methodology introduces a rigorous framework for propagating measurement uncertainty into PBLH diagnostics. By combining the simulation‑smoothing capabilities of the state‑space model with a Monte Carlo approach, it provides confidence intervals for each PBLH retrieval method. This quantification reveals meaningful differences in how individual diagnostics respond to varying atmospheric conditions. Overall, the framework represents a substantive step toward a more principled and transparent evaluation of PBLH retrievals.

Both the example profiles and the aggregated analyses show that explicit uncertainty quantification  enhances confidence in the PBLH retrieval and provide a clear message in the conistency across different approaches.

Some preliminary results also suggest that the revised MC estimation may be 
more robust to certain sources of error that can affect the standard plug-in 
application, particularly the gradient-based methods. This is due to the probabilistic 
treatment able to mitigate the sensitivity of PBLH retrieval to noisy or weak gradients, yielding 
more stable distributions, while the standard approach tends to produce erratic values.

Further analysis and validation efforts will strengthen the present 
work and clarify the practical implications of the revised MC estimation and 
its associated uncertainty quantification. Such follow-up studies will help 
determine how the probabilistic framework performs under a wider range of 
atmospheric conditions and establish its reliability relative to the 
standard plug-in approach.

Such progression is not straightforward, owing both to the absence of a 
definitive ground-truth reference for PBLH and to the inherently complex and 
multifaceted nature of the boundary layer itself. Both the literature and the 
present analysis highlight how different methods have the ability to capture distinct features of 
the boundary-layer structure, helping to clarify specific dynamical processes in different atmospheric  and sunlight condition.

Next working steps will include:
\begin{itemize}
    \item A detailed discussion of the Kalman smoothing results and the Maximum 
    Likelihood Estimation, evaluated over a broader set of profiles. This will 
    allow exploration of additional modelling and estimation choices and help 
    prevent potential artefacts associated with oversmoothing.

    \item A systematic analysis of the impact of the measurement uncertainty and the MC–procedure parameters, 
    such as the number of realizations $M$, the diagnostic variable function 
    $\bar{f}$ and the actual choices for $\mathbf{h}^{MC}$ and $u(\mathbf{h}^{MC})$, to quantify their influence on the resulting distribution and 
    estimation.

    \item An extension of the profile dataset, combined with a quantitative 
    classification of PBL regimes (Convective, Neutral, Stable), to assess how the revised 
    estimation behaves under different boundary–layer conditions across the full diurnal cycle, including daytime, nighttime, and twilight transitions..

    \item The development of an estimator or proxy for the uncertainty associated 
    with standard plug–in PBLH retrievals, enabling a direct comparison with the 
    proposed MC approach.

    \item The identification of quantitative metrics to evaluate the performance 
    of different methodologies in a variety of atmospheric scenarios.
    
\end{itemize}

\refstepcounter{section}
\section*{Code Availability}
\addcontentsline{toc}{section}{Code Availability}
\label{sec:code}

All code used for the state-space modelling, Kalman smoothing, Monte-Carlo simulation, and PBLH diagnostics is available at the following repository:

\begin{itemize}
    \item \url{https://github.com/TommasoLocatelli/GRUAN_EDA}
\end{itemize}

The analysis was performed in Python~3.11 using the libraries
\texttt{NumPy}, \texttt{Pandas}, \texttt{Statsmodels}, and the ongoing from-scratch development package \texttt{GRUANpy}.
The full workflow, including preprocessing, model fitting, and uncertainty quantification, is fully reproducible using the scripts provided.

\refstepcounter{section}
\section*{Data Availability}
\addcontentsline{toc}{section}{Data Availability}
\label{sec:data}

The radiosonde profiles used in this study are part of the GRUAN data record
and are publicly available through the GRUAN data portal
(\url{https://www.gruan.org}).

\refstepcounter{section}
\section*{AI Use Disclosure}
\addcontentsline{toc}{section}{AI Use Disclosure}
\label{sec:AI}

Artificial intelligence tools were used to support 
manuscript preparation, including assistance with text editing and literature discovery. All scientific analyses, 
state-space modelling, Kalman smoothing, Monte-Carlo simulations, 
and PBLH diagnostics were performed entirely by the authors. 
No AI system was used to generate, modify, or validate research 
results, figures, or data products. All AI-assisted text was 
reviewed, edited, and verified by the authors to ensure accuracy 
and consistency with the scientific content of the study.

\bibliography{ref}

\newpage
\refstepcounter{section}
\section*{Tables}
\addcontentsline{toc}{section}{Table List}
\label{sec:tbls}
\listoftables

\begin{table}[h!]
\centering
\hspace*{-0.8cm}
\resizebox{1.1\textwidth}{!}{%
\footnotesize
\begin{tabular}{l l r r r r r r}
\toprule
\textbf{Method} & \textbf{Tod} &
\textbf{Std Median} & \textbf{Std IQR} &
\textbf{MC Median} & \textbf{MC IQR} &
\textbf{Unc. Median} & \textbf{Unc. IQR} \\
\midrule
PM  & daytime   & 264.10 & 1078.44 & 758.48 & 1102.72 & 75.27 & 184.24 \\
PM  & twilight  & 117.87 & 7.90    & 117.58 & 9.04    & 10.24 & 2.77   \\
PM  & nighttime & 115.97 & 8.35    & 117.25 & 8.19    & 10.67 & 13.21  \\
\midrule
$\theta_v$ & daytime   & 1170.25 & 2045.20 & 1800.23 & 1670.12 & 74.04 & 826.72 \\
$\theta_v$ & twilight  & 410.67  & 1880.70 & 151.18  & 577.09  & 47.06 & 39.24  \\
$\theta_v$ & nighttime & 226.43  & 785.90  & 255.09  & 1132.53 & 47.95 & 122.72 \\
\midrule
RH  & daytime   & 1567.32 & 1966.54 & 1972.07 & 1358.99 & 44.02 & 856.31 \\
RH  & twilight  & 2190.49 & 1728.16 & 2561.95 & 1652.27 & 153.94 & 2004.64 \\
RH  & nighttime & 1264.02 & 2196.89 & 1739.42 & 1709.44 & 41.55 & 1016.98 \\
\midrule
RM  & daytime   & 1020.16 & 1026.69 & 998.62  & 956.47  & 62.86 & 75.89  \\
RM  & twilight  & 148.44  & 192.63  & 163.48  & 220.40  & 35.99 & 38.48  \\
RM  & nighttime & 236.14  & 376.14  & 248.79  & 434.81  & 47.40 & 58.43  \\
\bottomrule
\end{tabular}
}
\caption{Median and IQR of standard PBLH method application (Std), Monte Carlo estimate PBLH (MC), and uncertainty width (Unc.) for the Lindenberg dataset by time of the day. All values are expressed in meters [m].}
\label{tbl:lin_tod}
\end{table}

\begin{table}[ht]
\centering
\hspace*{-0.8cm}
\resizebox{1.1\textwidth}{!}{%
\footnotesize
\begin{tabular}{l l r r r r r r}
\toprule
\textbf{Method} & \textbf{Tod} &
\textbf{Std Median} & \textbf{Std IQR} &
\textbf{MC Median} & \textbf{MC IQR} &
\textbf{Unc. Median} & \textbf{Unc. IQR} \\
\midrule
PM  & daytime   & 82.64  & 395.85  & 286.82 & 417.91 & 133.08 & 265.86 \\
PM  & twilight  & 83.64  & 295.26  & 176.56 & 565.13 & 166.86 & 417.72 \\
PM  & nighttime & 259.99 & 506.27  & 266.52 & 476.03 & 101.25 & 280.30 \\
\midrule
$\theta_v$ & daytime   & 281.94 & 1501.01 & 1736.11 & 1643.27 & 156.29 & 1412.32 \\
$\theta_v$ & twilight  & 166.52 & 733.68  & 1961.92 & 1846.59 & 1361.56 & 2020.71 \\
$\theta_v$ & nighttime & 593.10 & 1668.70 & 1788.03 & 1376.19 & 107.72 & 1335.40 \\
\midrule
RH  & daytime   & 602.08 & 2270.45 & 2259.94 & 1466.55 & 345.91 & 1519.88 \\
RH  & twilight  & 677.67 & 1895.33 & 1795.62 & 1750.64 & 1202.75 & 1820.07 \\
RH  & nighttime & 1536.26 & 2473.89 & 2309.59 & 1254.24 & 69.61 & 1126.16 \\
\midrule
RM  & daytime   & 495.11 & 606.25  & 535.19 & 411.46 & 90.68 & 144.76 \\
RM  & twilight  & 494.96 & 679.27  & 665.35 & 384.46 & 95.90 & 147.03 \\
RM  & nighttime & 556.88 & 575.60  & 536.74 & 446.33 & 77.20 & 131.12 \\
\bottomrule
\end{tabular}
}
\caption{Median and IQR of standard PBLH method application (Std), Monte Carlo estimate PBLH (MC), and uncertainty width (Unc.) for the Hong Kong dataset by time of the day. All values are expressed in meters [m].}
\label{tbl:hko_tod}
\end{table}

\begin{table}[ht]
\centering
\hspace*{-0.8cm}
\resizebox{1.1\textwidth}{!}{%
\footnotesize
\begin{tabular}{l l r r r r r r}
\toprule
\textbf{Method} & \textbf{Tod} &
\textbf{Std Median} & \textbf{Std IQR} &
\textbf{MC Median} & \textbf{MC IQR} &
\textbf{Unc. Median} & \textbf{Unc. IQR} \\
\midrule
PM  & daytime   & 53.60  & 353.98  & 85.68  & 426.60  & 84.81  & 176.31 \\
PM  & nighttime & 6.80   & 11.10   & 8.95   & 15.80   & 10.75  & 39.19  \\
\midrule
$\theta_v$ & daytime   & 137.66 & 1540.07 & 1518.97 & 1786.77 & 69.82  & 741.73 \\
$\theta_v$ & nighttime & 36.10  & 309.11  & 546.15  & 1888.35 & 65.71  & 1192.05 \\
\midrule
RH  & daytime   & 724.23 & 1968.87 & 1818.86 & 1419.36 & 45.57  & 593.89 \\
RH  & nighttime & 190.41 & 1793.14 & 1848.97 & 1634.02 & 49.81  & 726.22 \\
\midrule
RM  & daytime   & 653.49 & 865.78  & 610.59  & 826.26  & 66.02  & 73.16  \\
RM  & nighttime & 237.51 & 601.23  & 277.12  & 637.23  & 55.24  & 59.80  \\
\bottomrule
\end{tabular}
}
\caption{Median and IQR of standard PBLH method application (Std), Monte Carlo estimate PBLH (MC), and uncertainty width (Unc.) for the Lauder dataset by time of the day. All values are expressed in meters [m].}
\label{tbl:lau_tod}
\end{table}

\begin{table}[ht]
\centering
\hspace*{-0.8cm}
\resizebox{1.1\textwidth}{!}{%
\footnotesize
\begin{tabular}{l l r r r r r r}
\toprule
\textbf{Method} & \textbf{Season} & \textbf{Std Med} & \textbf{Std IQR} & \textbf{MC Median} & \textbf{MC IQR} & \textbf{Unc. Median} & \textbf{Unc. IQR} \\
\midrule
PM  & winter & 118.07 & 18.61 & 121.12 & 145.37 & 54.80 & 124.32 \\
PM  & spring & 120.47 & 535.36 & 178.31 & 881.03 & 20.63 & 96.06 \\
PM  & summer & 121.42 & 1097.87 & 327.61 & 1288.95 & 29.99 & 80.84 \\
PM  & autumn & 118.22 & 26.42 & 124.21 & 515.58 & 20.29 & 76.54 \\
\hline
$\theta_v$ & winter & 407.57 & 1343.41 & 952.51 & 1438.01 & 60.01 & 188.14 \\
$\theta_v$ & spring & 591.61 & 1847.20 & 1264.29 & 2156.72 & 54.81 & 183.06 \\
$\theta_v$ & summer & 1022.11 & 2163.43 & 1708.39 & 2447.62 & 67.61 & 822.55 \\
$\theta_v$ & autumn & 451.84 & 1252.74 & 839.86 & 1336.50 & 74.13 & 225.53 \\
\hline
RH & winter & 1131.84 & 1875.46 & 1619.83 & 1413.92 & 39.59 & 399.88 \\
RH & spring & 1560.57 & 1980.52 & 1887.85 & 1561.56 & 47.99 & 1179.15 \\
RH & summer & 2024.94 & 1670.45 & 2266.23 & 1334.23 & 47.46 & 980.55 \\
RH & autumn & 984.05 & 2231.19 & 1829.72 & 1727.93 & 42.31 & 963.81 \\
\hline
RM & winter & 500.82 & 633.52 & 503.41 & 621.58 & 57.06 & 63.61 \\
RM & spring & 610.41 & 1128.23 & 614.37 & 1066.50 & 54.53 & 79.61 \\
RM & summer & 460.10 & 1398.76 & 586.54 & 1366.93 & 51.85 & 64.31 \\
RM & autumn & 544.99 & 756.73 & 595.15 & 629.84 & 46.76 & 63.42 \\
\hline
\end{tabular}
}
\caption{Median and IQR of standard PBLH method application (Std), Monte Carlo estimate PBLH (MC), and uncertainty width (Unc.) for the Lindenberg dataset by season. All values are expressed in meters [m].}
\label{tbl:lin_season_stats}
\end{table}

\begin{table}[ht]
\centering
\hspace*{-0.8cm}
\resizebox{1.1\textwidth}{!}{%
\footnotesize
\begin{tabular}{l l r r r r r r}
\toprule
\textbf{Method} & \textbf{Season} & \textbf{Std Med} & \textbf{Std IQR} & \textbf{MC Median} & \textbf{MC IQR} & \textbf{Unc. Median} & \textbf{Unc. IQR} \\
\midrule
PM  & winter & 314.58 & 479.13 & 292.42 & 452.25 & 60.51 & 110.64 \\
PM  & spring & 270.57 & 456.68 & 360.49 & 385.65 & 133.08 & 275.67 \\
PM  & summer & 78.75  & 187.53 & 166.76 & 404.53 & 176.50 & 342.09 \\
PM  & autumn & 85.63  & 410.27 & 217.88 & 410.31 & 191.29 & 369.23 \\
\hline
$\theta_v$ & winter & 404.19 & 1568.24 & 1700.77 & 1602.61 & 51.19 & 720.08 \\
$\theta_v$ & spring & 328.05 & 1529.69 & 1632.17 & 1274.79 & 92.20 & 1370.69 \\
$\theta_v$ & summer & 288.73 & 1150.44 & 2124.59 & 1879.77 & 982.47 & 1987.46 \\
$\theta_v$ & autumn & 412.88 & 1530.14 & 1847.29 & 1476.86 & 287.52 & 1369.21 \\
\hline
RH & winter & 1351.20 & 2378.58 & 2415.99 & 1347.13 & 47.89 & 950.21 \\
RH & spring & 877.88  & 2108.33 & 1922.35 & 1329.15 & 171.06 & 1422.66 \\
RH & summer & 597.99  & 2098.70 & 2076.82 & 1758.36 & 1040.29 & 1999.69 \\
RH & autumn & 963.59  & 2536.12 & 2344.99 & 1306.57 & 135.13 & 1265.79 \\
\hline
RM & winter & 465.57 & 587.69 & 500.84 & 393.01 & 64.91 & 94.44 \\
RM & spring & 587.71 & 476.10 & 568.73 & 399.83 & 67.69 & 91.08 \\
RM & summer & 462.67 & 671.11 & 620.41 & 444.11 & 132.53 & 194.02 \\
RM & autumn & 531.83 & 652.00 & 598.13 & 448.86 & 111.16 & 199.07 \\
\hline
\end{tabular}
}
\caption{Median and IQR of standard PBLH method application (Std), Monte Carlo estimate PBLH (MC), and uncertainty width (Unc.) for the Hong Kong dataset by season. All values are expressed in meters [m].}
\label{tbl:hko_season_stats}
\end{table}

\begin{table}[ht]
\centering
\hspace*{-0.8cm}
\resizebox{1.1\textwidth}{!}{%
\footnotesize
\begin{tabular}{l l r r r r r r}
\toprule
\textbf{Method} & \textbf{Season} & \textbf{Std Med} & \textbf{Std IQR} & \textbf{MC Median} & \textbf{MC IQR} & \textbf{Unc. Median} & \textbf{Unc. IQR} \\
\midrule
PM  & winter & 16.70 & 255.31 & 31.63 & 528.87 & 70.50 & 260.55 \\
PM  & spring & 8.50  & 48.60  & 9.63  & 109.07 & 21.33 & 89.36 \\
PM  & summer & 10.10 & 25.00  & 10.77 & 20.63  & 12.06 & 40.71 \\
PM  & autumn & 10.00 & 97.16  & 14.35 & 403.66 & 33.25 & 132.64 \\
\hline
$\theta_v$ & winter & 142.96 & 1216.54 & 1222.78 & 1697.85 & 98.00 & 1131.41 \\
$\theta_v$ & spring & 112.81 & 1442.89 & 1462.02 & 1974.10 & 58.42 & 966.10 \\
$\theta_v$ & summer & 85.61  & 188.51  & 145.73 & 1564.61 & 49.04 & 414.24 \\
$\theta_v$ & autumn & 33.10  & 586.59  & 1493.37 & 2033.52 & 90.08 & 1347.35 \\
\hline
RH & winter & 515.15 & 1969.72 & 2043.15 & 1678.92 & 112.23 & 714.81 \\
RH & spring & 1006.93 & 1833.82 & 1831.96 & 918.90  & 38.98 & 1013.92 \\
RH & summer & 191.41 & 2003.73 & 1446.65 & 1959.08 & 36.65 & 575.54 \\
RH & autumn & 154.41 & 1805.73 & 2015.45 & 1546.40 & 51.40 & 580.27 \\
\hline
RM & winter & 567.85 & 797.86 & 661.09 & 762.14 & 73.33 & 72.59 \\
RM & spring & 501.15 & 711.36 & 392.33 & 699.68 & 73.67 & 72.83 \\
RM & summer & 168.06 & 409.83 & 151.40 & 484.18 & 47.23 & 47.36 \\
RM & autumn & 592.64 & 740.74 & 589.93 & 748.99 & 66.05 & 78.26 \\
\hline
\end{tabular}
}
\caption{Median and IQR of standard PBLH method application (Std), Monte Carlo estimate PBLH (MC), and uncertainty width (Unc.) for the Lauder dataset by season. All values are expressed in meters [m].}
\label{tbl:lau_season_stats}
\end{table}

\end{document}